\documentclass[11pt]{article}
 
\usepackage{times}

\usepackage{color}
\definecolor{indigo}{RGB}{0,0,120}
\usepackage[colorlinks=true, linkcolor=indigo, citecolor=blue, urlcolor=indigo]{hyperref}

\usepackage{amsmath,amssymb}

\usepackage{graphicx}

\usepackage{subcaption} % for figures to appear in two columns

\usepackage{mdframed}

\def\imply{\Rightarrow}

\newcommand{\tl}[1]{\tilde{#1}}
\newcommand{\dd}[2]{\frac {\partial #1}{\partial #2}}

\newcommand{\pdr}{\partial}
\newcommand{\DD}[2]{\frac {d #1}{d #2}}
\newcommand{\grad}{{\bf \nabla}}

\newcommand{\beq}{\begin{equation}}
\newcommand{\eeq}{\end{equation}}
\newcommand{\beqs}{\begin{eqnarray}}
\newcommand{\eeqs}{\end{eqnarray}}

\newcommand{\half}{\frac{1}{2}}
\newcommand{\ov}[1]{\frac{1}{#1}}
\newcommand{\fr}[2]{\frac{#1}{#2}}

\def\al{\alpha}		 		 
\def\del{\delta}			\def\eps{\epsilon} 
			 
\def\la{\lambda}				
		\def\tht{\theta}	
		\def\om{\omega}		\def\Om{\Omega}

\usepackage{titlesec}
\titleformat{\section}{\normalsize\bfseries}{\thesection}{1em}{}
\titleformat{\subsection}{\small\bfseries}{\thesubsection}{1em}{}
\titleformat{\subsubsection}{\small\bfseries}{\thesubsubsection}{1em}{}

\usepackage{calligra} % for script and cursive characters like script r
\DeclareMathAlphabet{\mathcalligra}{T1}{calligra}{m}{n}
\DeclareFontShape{T1}{calligra}{m}{n}{<->s*[2.2]callig15}{}

\newcount\colveccount  % column vec \colvec{nrows}{a11 & a12}{a21 & a22}
\newcommand*\colvec[1]{\global\colveccount#1  \begin{pmatrix} \colvecnext} \def\colvecnext#1{#1 \global\advance\colveccount-1
        \ifnum\colveccount>0 \\ \expandafter\colvecnext
        \else \end{pmatrix} \fi}

\newcommand{\bfb}{{\bf b}}
\newcommand{\bfc}{{\bf c}}

\newcommand{\bfr}{{\bf r}}

\newcommand{\bfp}{{\bf p}}
\newcommand{\bfq}{{\bf q}}
\newcommand{\bfA}{{\bf A}}

\newcommand{\bfF}{{\bf F}}
\newcommand{\bfR}{{\bf R}}
\newcommand{\bfL}{{\bf L}}
\newcommand{\bfP}{{\bf P}}
\newcommand{\bfQ}{{\bf Q}}
\newcommand{\bfJ}{{\bf J}}

\newcommand{\mR}{{\mathbb{R}}}
\newcommand{\mS}{{\mathbb{S}}}

\begin{document}

%-------------------------------------

% \thispagestyle{empty}

\title{\normalsize 
% \hfill {\tt arXiv:} \\
\vskip 0.1mm \Large An introduction to the classical three-body problem \\ \vskip 0.6mm
\normalsize From periodic solutions to instabilities and chaos}
\author{{\sc Govind S. Krishnaswami and Himalaya Senapati}
\\ \small
Chennai Mathematical Institute,  SIPCOT IT Park, Siruseri 603103, India
\\ \small
 Email: {\tt govind@cmi.ac.in, himalay@cmi.ac.in}}

\date{January 22, 2019 \\ \vspace{.2cm} Published in \href{https://www.ias.ac.in/article/fulltext/reso/024/01/0087-0114}{Resonance} {\bf 24}(1), 87-114, January (2019)}

\maketitle

\begin{abstract} \normalsize

The classical three-body problem arose in an attempt to understand the effect of the Sun on the Moon's Keplerian orbit around the Earth. It has attracted the attention of some of the best physicists and mathematicians and led to the discovery of chaos. We survey the three-body problem in its historical context and use it to introduce several ideas and techniques  that have been developed to understand classical mechanical systems.

\end{abstract} \normalsize


\small

\tableofcontents

\normalsize

%--------------
\section{Introduction}
%--------------

The three-body problem is one of the oldest problems in classical dynamics that continues to throw up surprises. It has challenged scientists from Newton's time to the present. It arose in an attempt to understand the Sun's effect on the motion of the Moon around the Earth. This was of much practical importance in marine navigation, where lunar tables were necessary to accurately determine longitude at sea (see Box 1). The study of the three-body problem led to the discovery of the planet Neptune (see Box 2), it explains the location and stability of the Trojan asteroids and has furthered our understanding of the stability of the solar system \cite{laskar}. Quantum mechanical variants of the three-body problem are relevant to the Helium atom and water molecule \cite{gutzwiller-book}.

\begin{center}
	\begin{mdframed}
{\bf Box 1:} The {\bf Longitude Act} (1714) of the British Parliament offered \pounds 20,000 for a method to determine the longitude at sea to an accuracy of half a degree. This was important for marine navigation at a time of exploration of the continents. In the absence of accurate clocks that could function at sea, a lunar table along with the observed position of the Moon was the principal method of estimating the longitude. Leonhard Euler\footnote{Euler had gone blind when he developed much of his lunar theory!}, Alexis Clairaut and Jean-Baptiste d'Alembert competed to develop a theory accounting for solar perturbations to the motion of the Moon around the Earth. For a delightful account of this chapter in the history of the three-body problem, including Clairaut's explanation of the annual $40^\circ$ rotation of the lunar perigee (which had eluded Newton), see \cite{bodenmann-lunar-battle}. Interestingly, Clairaut's use of Fourier series in the three-body problem (1754) predates their use by Joseph Fourier in the analysis of heat conduction!
	\end{mdframed}
\end{center}

\begin{center}
	\begin{mdframed}
{\bf Box 2: Discovery of Neptune:} The French mathematical astronomer Urbain Le Verrier (1846) was intrigued by discrepancies between the observed and Keplerian orbits of Mercury and Uranus. He predicted the existence of Neptune (as was widely suspected) and calculated its expected position based on its effects on the motion of Uranus around the Sun (the existence and location of Neptune was independently inferred by John Adams in Britain). The German astronomer Johann Galle (working with his graduate student Heinrich d'Arrest) discovered Neptune within a degree of Le Verrier's predicted position on the very night that he received the latter's letter. It turned out that both Adams' and Le Verrier's heroic calculations were based on incorrect assumptions about Neptune, they were extremely lucky to stumble upon the correct location!
	\end{mdframed}
\end{center}

The three-body problem admits many `regular' solutions such as the collinear and equilateral periodic solutions of Euler and Lagrange as well as the more recently discovered figure-8 solution. On the other hand, it can also display chaos as serendipitously discovered by Poincar\'e. Though a general solution in closed form is not known, Sundman while studying binary collisions, discovered an exceptionally slowly converging series representation of solutions in fractional powers of time.

The importance of the three-body problem goes beyond its application to the motion of celestial bodies. As we will see, attempts to understand its dynamics have led to the discovery of many phenomena (e.g., abundance of periodic motions, resonances (see Box 3), homoclinic points, collisional and non-collisional singularities, chaos and KAM tori) and techniques (e.g., Fourier series, perturbation theory, canonical transformations and regularization of singularities) with applications across the sciences. The three-body problem provides a context in which to study the development of classical dynamics as well as a window into several areas of mathematics (geometry, calculus and dynamical systems).

\begin{center}
	\begin{mdframed}
{\bf Box 3: Orbital resonances:} The simplest example of an orbital resonance occurs when the periods of two orbiting bodies (e.g., Jupiter and Saturn around the Sun) are in a ratio of small whole numbers ($T_S/T_J \approx 5/2$). Resonances can enhance their gravitational interaction and have both stabilizing and destabilizing effects. For instance, the moons Ganymede, Europa and Io are in a stable $1:2:4$ orbital resonance around Jupiter. The Kirkwood gaps in the asteroid belt are probably due to the destabilizing resonances with Jupiter. Resonances among the natural frequencies of a system (e.g., Keplerian orbits of a pair of moons of a planet) often lead to difficulties in naive estimates of the effect of a perturbation (say of the moons on each other).
	\end{mdframed}
\end{center}

%--------------
\section{Review of the Kepler problem}
%--------------

\begin{figure}[h] 
\center
\includegraphics[width=8cm]{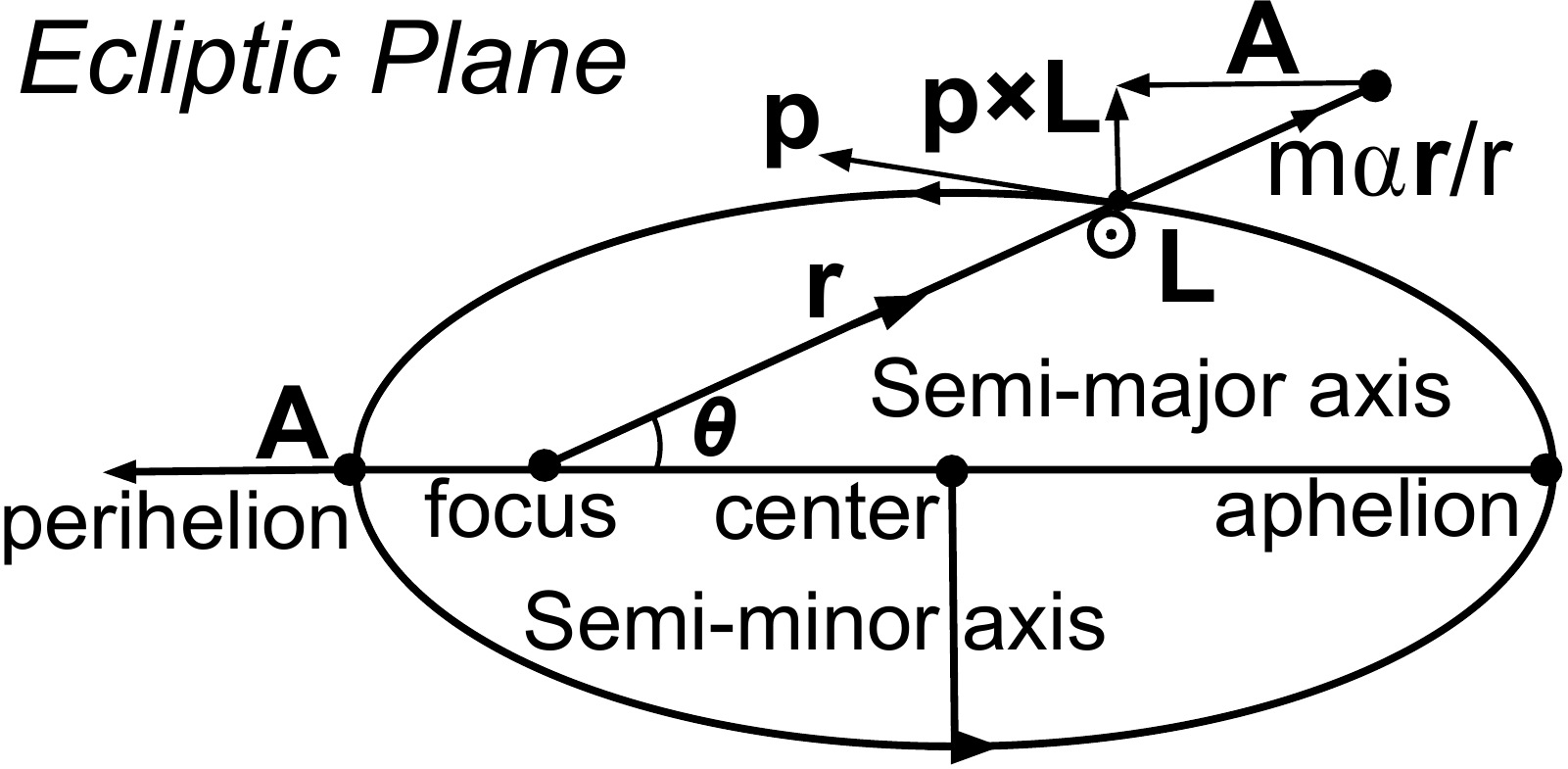}
\caption{\footnotesize Keplerian ellipse in the ecliptic plane of motion showing the constant LRL vector $\bfA$. The constant angular momentum $\bfL$ points out of the ecliptic plane.}
\label{f:lrl-vector}
\end{figure}

As preparation for the three-body problem, we begin by reviewing some key features of the two-body problem. If we ignore the non-zero size of celestial bodies, Newton's second law for the motion of two gravitating masses states that
	\beq
	m_1 \ddot \bfr_1 = \al \frac{ (\bfr_2 - \bfr_1)}{|\bfr_1 - \bfr_2|^3}  \quad \text{and} \quad 
	m_2 \ddot \bfr_2 = \al \frac{ (\bfr_1 - \bfr_2)}{|\bfr_1 - \bfr_2|^3}.
	\label{e:two-body-newton-ode}
	\eeq
Here, $\al = G m_1 m_2$ measures the strength of the gravitational attraction and dots denote time derivatives. This system has six degrees of freedom, say the three Cartesian coordinates of each mass $\bfr_1 = (x_1,y_1,z_1)$ and $\bfr_2 = (x_2,y_2,z_2)$. Thus, we have a system of 6 nonlinear (due to division by $|\bfr_1-\bfr_2|^3$), second-order ordinary differential equations (ODEs) for the positions of the two masses. It is convenient to switch from $\bfr_1$ and $\bfr_2$ to the center of mass (CM) and relative coordinates
	\beq
	\bfR = \frac{m_1 \bfr_1 + m_2 \bfr_2}{m_1 + m_2} \quad \text{and} \quad
	\bfr = \bfr_2 - \bfr_1.
	\eeq
In terms of these, the equations of motion become
	\beq
	M \ddot \bfR = 0 \quad \text{and} \quad m \ddot \bfr = - \frac{\al}{|\bfr|^3} \bfr.
	\eeq
Here, $M = m_1 + m_2$ is the total mass and $m = m_1 m_2/M$ the `reduced' mass. An advantage of these variables is that in the absence of external forces the CM moves at constant velocity, which can be chosen to vanish by going to a frame moving with the CM. The motion of the relative coordinate $\bfr$ decouples from that of $\bfR$ and describes a system with three degrees of freedom $\bfr = (x,y,z)$. Expressing the conservative gravitational force in terms of the gravitational potential $V = - \alpha/|\bfr|$, the equation for the relative coordinate $\bfr$ becomes
	\beq
	\dot \bfp \equiv m \ddot \bfr = - \grad_\bfr V = - \left(\dd{V}{x}, \dd{V}{y}, \dd{V}{z} \right)
	\eeq
where $\bfp = m \dot \bfr$ is the relative momentum. Taking the dot product with the `integrating factor' $\dot \bfr = (\dot x, \dot y, \dot z)$, we get 
	\beq
	m \dot \bfr \cdot \ddot \bfr = \fr{d}{dt}\left(\half m \dot \bfr^2 \right) = - \left( \dd{V}{x} \; \dot x + \dd{V}{y} \; \dot y + \dd{V}{z} \; \dot z \right) =  - \fr{dV}{dt},
	\label{e:energy-kepler-cm-frame}
	\eeq
which implies that the energy $E \equiv \half m \dot \bfr^2 + V$ or Hamiltonian $\frac{\bfp^2}{2m} + V$ is conserved. The relative angular momentum ${\bf L} = \bfr \times m \dot \bfr = \bfr \times \bfp$ is another constant of motion as the force is central\footnote{The conservation of angular momentum in a central force is a consequence of rotation invariance: $V = V(|\bfr|)$ is independent of polar and azimuthal angles. More generally, Noether's theorem relates continuous symmetries to conserved quantities.}: $\dot \bfL = \dot \bfr \times \bfp + \bfr \times \dot \bfp = 0 + 0$. The constancy of the direction of $\bfL$ implies planar motion in the CM frame: $\bfr$ and $\bfp$ always lie in the `ecliptic plane' perpendicular to $\bfL$, which we take to be the $x$-$y$ plane with origin at the CM (see Fig.~\ref{f:lrl-vector}). The Kepler problem is most easily analyzed in plane-polar coordinates $\bfr = (r, \tht)$ in which the energy $E = \half m \dot r^2 + V_{\rm eff}(r)$ is the sum of a radial kinetic energy and an effective potential energy $V_{\rm eff} = \fr{L_z^2}{2 m r^2} + V(r)$. Here, $L_z = m r^2 \dot \tht$ is the vertical component of angular momentum and the first term in $V_{\rm eff}$ is the centrifugal `angular momentum barrier'. Since $\bfL$ (and therefore $L_z$) is conserved, $V_{\rm eff}$ depends only on $r$. Thus, $\tht$ does not appear in the Hamiltonian: it is a `cyclic' coordinate. Conservation of energy constrains $r$ to lie between `turning points', i.e., zeros of $E - V_{\rm eff}(r)$ where the radial velocity $\dot r$ momentarily vanishes. One finds that the orbits are Keplerian ellipses for $E < 0$ along with parabolae and hyperbolae for $E \geq 0$: $r(\tht) = \rho(1 + \eps \cos \tht)^{-1}$ \cite{goldstein,hand-finch}. Here, $\rho = L_z^2/m\al$ is the radius of the circular orbit corresponding to angular momentum $L_z$, $\eps$ the eccentricity and $E = - \frac{\al}{2\rho} (1 - \eps^2)$ the energy.

In addition to $E$ and $\bfL$, the Laplace-Runge-Lenz (LRL) vector ${\bf A} = \bfp \times {\bf L} - m \al \: \hat r$ is another constant of motion. It points along the semi-major axis from the CM to the perihelion and its magnitude determines the eccentricity of the orbit. Thus, we have $7$ conserved quantities: energy and three components each of $\bfL$ and $\bfA$. However, a system with three degrees of freedom has a six-dimensional phase space (space of coordinates and momenta, also called the state space) and if it is to admit continuous time evolution, it  cannot have more than 5 independent conserved quantities. The apparent paradox is resolved once we notice that $E$, $\bfL$ and $\bfA$ are not all independent; they satisfy two relations\footnote{Wolfgang Pauli (1926) derived the quantum mechanical spectrum of the Hydrogen atom using the relation between $E, \bfL^2$ and $\bfA^2$ before the development of the Schr\"odinger equation. Indeed, if we postulate circular Bohr orbits which have zero eccentricity ($\bfA = 0$) and quantized angular momentum $\bfL^2 = n^2 \hbar^2$, then $E_n = - \fr{m \al^2 }{2 \hbar^2 n^2}$ where $\al = e^2/4 \pi \epsilon_0$ is the electromagnetic analogue of $G m_1 m_2$.}:
	\beq
	\bfL \cdot \bfA = 0 \quad \text{and} \quad E = \frac{\bfA^2 - m^2 \alpha^2}{2 m \bfL^2}.
	\eeq
Newton used the solution of the two-body problem to understand the orbits of planets and comets. He then turned his attention to the motion of the Moon around the Earth. However, lunar motion is significantly affected by the Sun. For instance, $\bfA$ is {\it not} conserved and the lunar perigee rotates by $40^\circ$ per year. Thus, he was led to study the Moon-Earth-Sun three-body problem.

%--------------
\section{The three-body problem}
%--------------

We consider the problem of three point masses ($m_a$ with position vectors $\bfr_a$ for $a = 1,2,3$) moving under their mutual gravitational attraction. This system has 9 degrees of freedom, whose dynamics is determined by 9 coupled second order nonlinear ODEs:
	\beq 
	m_a \fr{d^2\bfr_a}{dt^2} = \sum_{b \neq a} G m_a m_b \fr{\bfr_b-\bfr_a}{|\bfr_b-\bfr_a |^3} \quad \text{for} \quad a = 1,2 \; \text{and} \; 3.
	\label{e:newtonian-3body-ODE}
	\eeq
As before, the three components of momentum $\bfP = \sum_a m_a \dot \bfr_a$, three components of angular momentum $\bfL = \sum_a \bfr_a \times \bfp_a$ and energy 
	\beq
	E = \half \sum_{a=1}^3 m_a \dot \bfr_a^2 - \sum_{a < b} \frac{G m_a m_b}{|\bfr_a - \bfr_b|}  \equiv T + V
	\eeq
furnish $7$ independent conserved quantities. Lagrange used these conserved quantities to reduce the above equations of motion to 7 first order ODEs (see Box 4).

\begin{center}
	\begin{mdframed}
	{\bf Box 4: Lagrange's reduction from 18 to 7 equations:} The 18 phase space variables of the 3-body problem (components of $\bfr_1, \bfr_2,  \bfr_3, \bfp_1, \bfp_2, \bfp_3$) satisfy 18 first order ordinary differential equations (ODEs) $\dot \bfr_a = \bfp_a$, $\dot \bfp_a = -\grad_{\bfr_a} V$. Lagrange (1772) used the conservation laws to reduce these ODEs to a system of 7 first order ODEs. Conservation of momentum determines 6 phase space variables comprising the location $\bfR_{\rm CM}$ and momentum $\bfP$ of the center of mass. Conservation of angular momentum $\bfL = \sum \bfr_a \times \bfp_a$ and energy $E$ lead to 4 additional constraints. By using one of the coordinates as a parameter along the orbit (in place of time), Lagrange reduced the three-body problem to a system of $7$ first order nonlinear ODEs.
	\end{mdframed}
\end{center}

\begin{figure}[h] 
\center
\includegraphics[width=6cm]{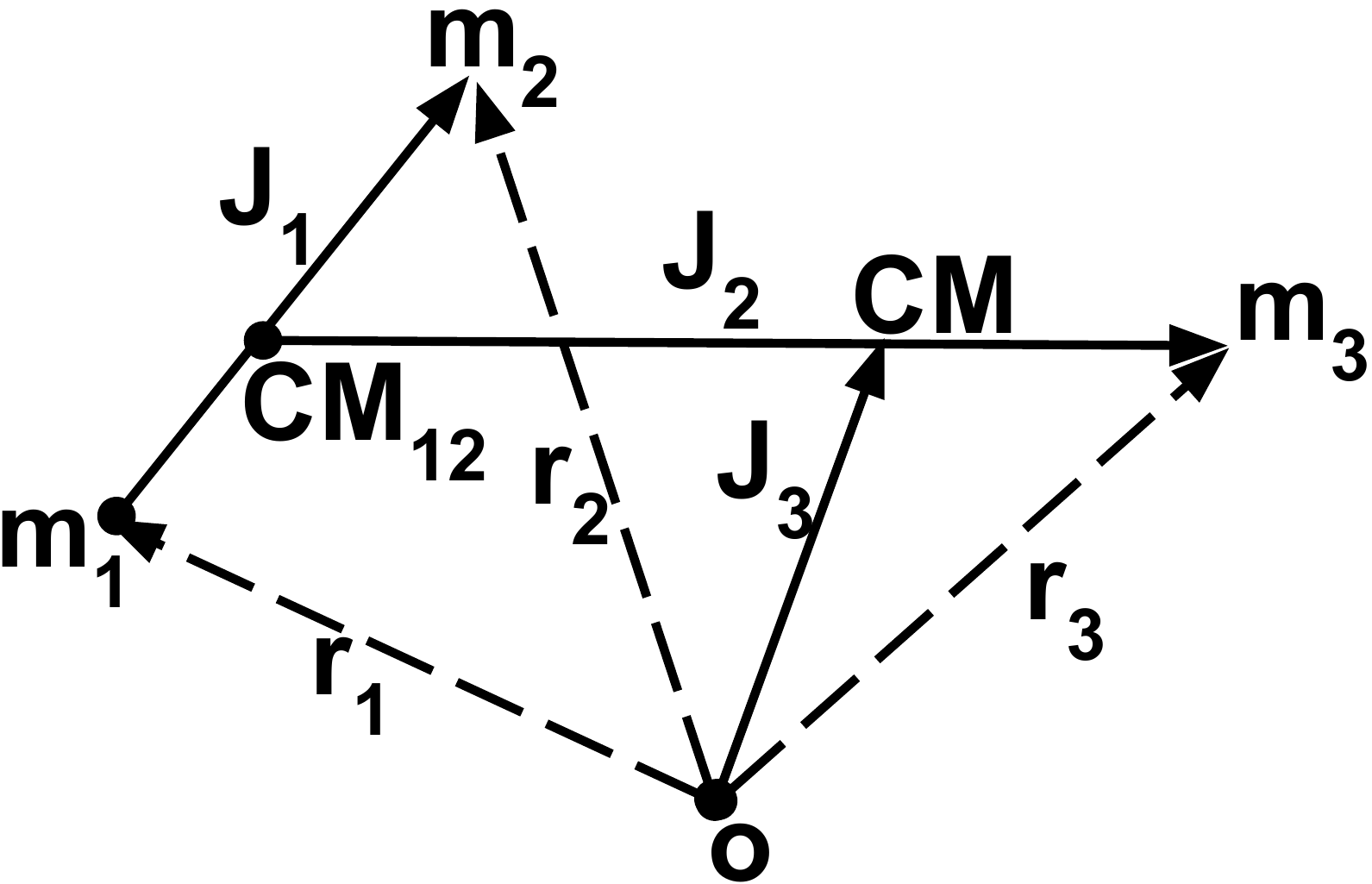}
\caption{\footnotesize Jacobi vectors $\bfJ_1, \bfJ_2$ and $\bfJ_3$ for the three-body problem. {\bf O} is the origin of the coordinate system while CM$_{12}$ is the center of mass of particles 1 and 2.}
\label{f:jacobi-coords}
\end{figure}

{\bf Jacobi vectors} (see Fig.~\ref{f:jacobi-coords}) generalize the notion of CM and relative coordinates to the 3-body problem \cite{Rajeev}. They are defined as 
	\beq
	\label{e:jacobi-coord}
	\bfJ_1 =  \bfr_2 - \bfr_1, \quad \bfJ_2 = \bfr_3 - \fr{m_1 \bfr_1 + m_2 \bfr_2}{m_1+m_2} \quad \text{and} \quad \bfJ_3 = \fr{m_1 \bfr_1 + m_2 \bfr_2 + m_3 \bfr_3}{m_1 + m_2 +m_3}.
	\eeq
$\bfJ_3$ is the coordinate of the CM, $\bfJ_1$ the position vector of $m_2$ relative to $m_1$ and $\bfJ_2$  that of $m_3$ relative to the CM of $m_1$ and $m_2$.  A nice feature of Jacobi vectors is that the kinetic energy $T = \half \sum_{a = 1,2,3} m_a \dot \bfr_a^2$ and moment of inertia $I = \sum_{a = 1,2,3} m_a \bfr_a^2$, regarded as quadratic forms, remain diagonal\footnote{A quadratic form $\sum_{a,b} r_a Q_{ab} r_b$ is diagonal if $Q_{ab} = 0$ for $a \ne b$.  Here, $\ov{M_1} = \ov{m_1} + \ov{m_2}$ is the reduced mass of the first pair, $\ov{M_2} = \ov{m_1+m_2}+\ov{m_3}$ is the reduced mass of $m_3$ and the ($m_1$, $m_2$) system and $M_3 = m_1 + m_2 + m_3$ the total mass.}:
	\beq
	\label{e:jacobi-coord-ke-mom-inertia}
	T = \half \sum_{1 \leq a \leq 3} M_a \dot \bfJ_a^2 \quad \text{and} \quad I = \sum_{1 \leq a \leq 3} M_a \bfJ_a^2.
	\eeq
What is more, just as the potential energy $- \al/|\bfr|$ in the two-body problem is a function only of the relative coordinate $\bfr$, here the potential energy $V$ may be expressed entirely in terms of $\bfJ_1$ and $\bfJ_2$:
	\beq
	V = - \frac{G m_1 m_2}{|\bfJ_1|} - \frac{G m_2 m_3}{|\bfJ_2 - \mu_1 \bfJ_1|} - \frac{G m_3 m_1}{|\bfJ_2 + \mu_2 \bfJ_1|} \quad \text{where} \quad \mu_{1,2} = \frac{m_{1,2}}{m_1 + m_2}.
	\label{e:jacobi-coord-potential}
	\eeq
Thus, the components of the CM vector $\bfJ_3$ are cyclic coordinates in the Hamiltonian $H = T + V$. In other words, the center of mass motion ($\ddot \bfJ_3 = 0$) decouples from that of $\bfJ_1$ and $\bfJ_2$.

An instantaneous configuration of the three bodies defines a triangle with masses at its vertices. The moment of inertia about the center of mass $I_{\rm CM} = M_1 \bfJ_1^2 + M_2 \bfJ_2^2$ determines the size of the triangle. For instance, particles suffer a triple collision when $I_{\rm CM} \to 0$ while $I_{\rm CM} \to \infty $ when one of the bodies flies off to infinity.

%------------------
\section{Euler and Lagrange periodic solutions}
%-------------------

\begin{figure}	
	\centering
	\begin{subfigure}[t]{3in}
		\centering
		\includegraphics[width=5cm]{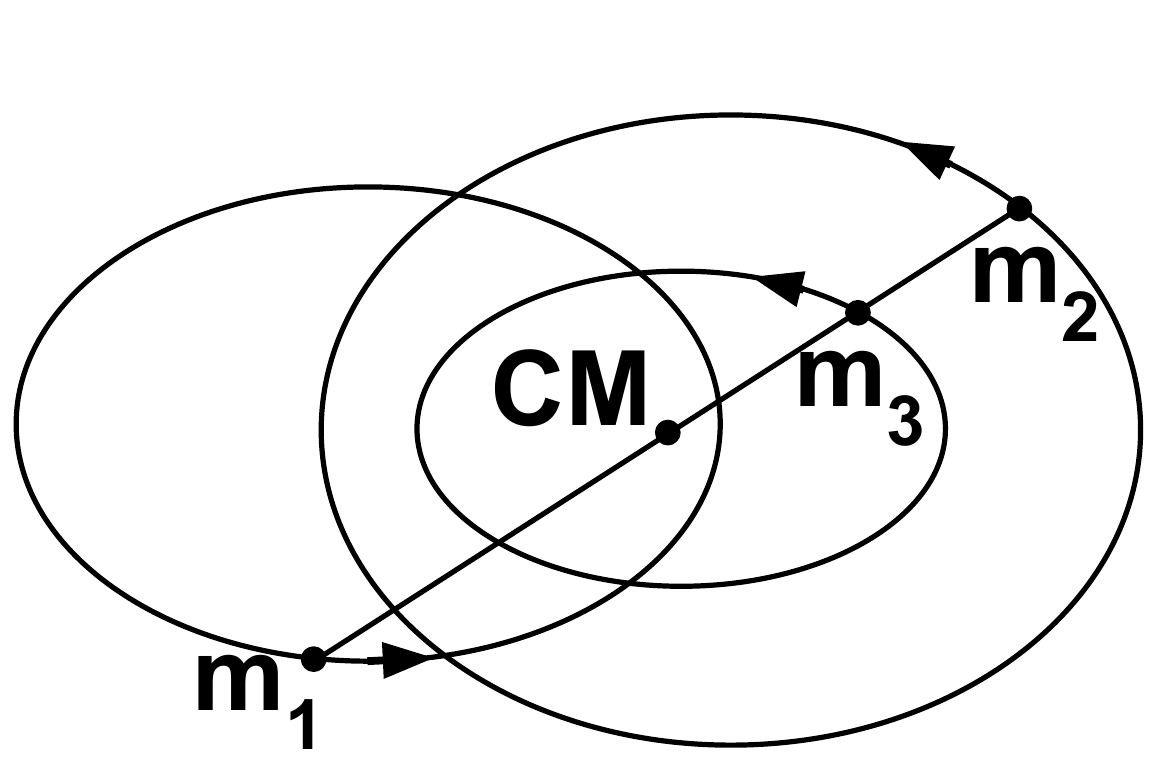}
		\caption{\footnotesize Masses traverse Keplerian ellipses with one focus at the CM.}
		\label{f:euler-periodic}		
	\end{subfigure}
	\quad
	\begin{subfigure}[t]{3in}
		\centering
		\includegraphics[width=3cm]{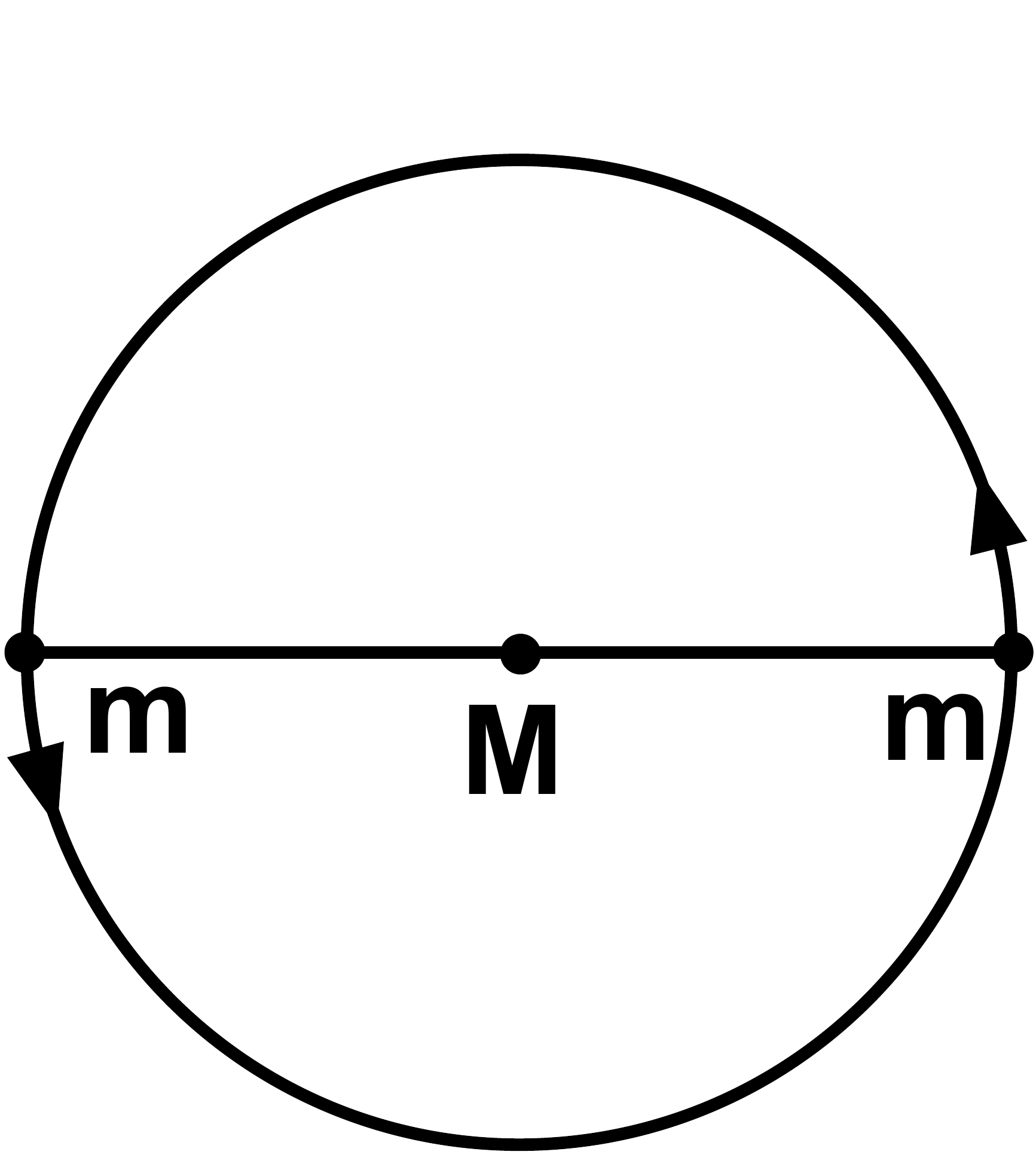}
		\caption{\footnotesize Two equal masses $m$ in a circular orbit around a third mass $M$ at their CM.}
		\label{f:euler-eq-mass}		
	\end{subfigure}	
	\caption{\footnotesize Euler collinear periodic solutions of the three-body problem. The constant ratios of separations are functions of the mass ratios alone.}
	\label{f:three-body-periodic}
\end{figure}

\begin{figure}	
	\centering
	\includegraphics[width=6cm]{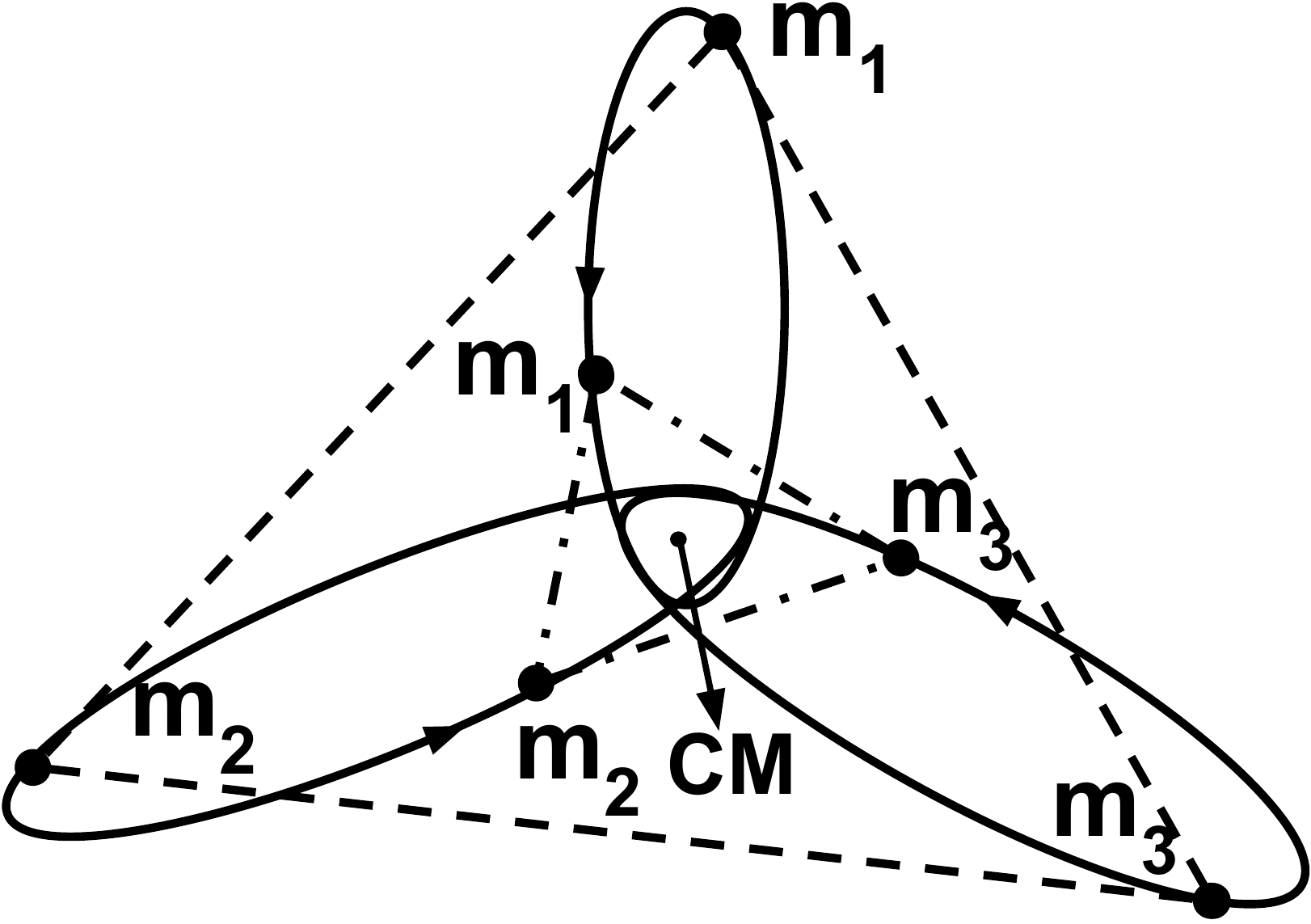}
	\caption{\footnotesize Lagrange's periodic solution with three bodies at vertices of equilateral triangles. The constant ratios of separations are functions of the mass ratios alone.}
	\label{f:lagrange-periodic}
\end{figure}

The planar three-body problem is the special case where the masses always lie on a fixed plane. For instance, this happens when the CM is at rest ($\dot \bfJ_3 = 0$) and the angular momentum about the CM vanishes ($\bfL_{\rm CM} = M_1 \bfJ_1 \times \dot \bfJ_1 + M_2 \bfJ_2 \times \dot \bfJ_2 = 0$). In 1767, the Swiss scientist Leonhard Euler discovered simple periodic solutions to the planar three-body problem where the masses are always collinear, with each body traversing a Keplerian orbit about their common CM. The line through the masses rotates about the CM with the ratio of separations remaining constant (see Fig.~\ref{f:euler-periodic}). The Italian/French mathematician Joseph-Louis Lagrange rediscovered Euler's solution in 1772 and also found new periodic solutions where the masses are always at the vertices of equilateral triangles whose size and angular orientation may change with time (see Fig.~\ref{f:lagrange-periodic}). In the limiting case of zero angular momentum, the three bodies move toward/away from their CM along straight lines. These implosion/explosion solutions are called Lagrange homotheties.

It is convenient to identify the plane of motion with the complex plane $\mathbb{C}$ and let the three complex numbers $z_{a=1,2,3}(t)$ denote the positions of the three masses at time $t$. E.g., the real and imaginary parts of $z_1$ denote the Cartesian components of the position vector $\bfr_1$ of the first mass. In Lagrange's solutions, $z_a(t)$ lie at vertices of an equilateral triangle while they are collinear in Euler's solutions. In both cases, the force on each body is always toward the common center of mass and proportional to the distance from it. For instance, the force on $m_1$ in a Lagrange solution is
	\beq
	\bfF_1 = G m_1 m_2 \fr{\bfr_2 - \bfr_1}{|\bfr_2 - \bfr_1|^3} + G m_1 m_3 \fr{\bfr_3 - \bfr_1}{|\bfr_3 - \bfr_1|^3} = \fr{Gm_1}{d^3} \left( m_1 \bfr_1 + m_2 \bfr_2 + m_3 \bfr_3 - M_3 \bfr_1 \right)
	\eeq
where $d = |\bfr_2 - \bfr_1| = |\bfr_3 - \bfr_1|$ is the side-length of the equilateral triangle and $M_3 = m_1 + m_2 + m_3$. Recalling that $\bfr_{\rm CM} = (m_1 \bfr_1 + m_2 \bfr_2 + m_3 \bfr_3)/M_3,$ we get
	\beq
	\bfF_1 = \fr{Gm_1}{d^3} M_3 \left( \bfr_{\rm CM} - \bfr_1 \right) \equiv G m_1 \delta_1 \fr{\bfr_{\rm CM} - \bfr_1}{|\bfr_{\rm CM} - \bfr_1|^3} 
	\eeq
where $\delta_1 = M_3 |\bfr_{\rm CM} - \bfr_1|^3/d^3$ is a function of the masses alone\footnote{Indeed, $\bfr_{\rm CM} - \bfr_1 = \left(m_2 (\bfr_2-\bfr_1) + m_3 (\bfr_3 - \bfr_1) \right)/M_3 \equiv \left( m_2 {\bf b} + m_3 {\bf c} \right)/ M_3$ where $\bfb$ and $\bfc$ are two of the sides of the equilateral triangle of length $d$. This leads to $|(\bfr_{\rm CM}-\bfr_1)/d| = \sqrt{m_2^2 + m_3^2 + m_2 m_3 }/M_3$ which is a function of masses alone. }. Thus, the equation of motion for $m_1$,
	\beq
	m_1 \ddot \bfr_1 = G m_1 \delta_1 \fr{\bfr_{\rm CM} - \bfr_1}{|\bfr_{\rm CM} - \bfr_1|^3},
	\eeq
takes the same form as in the two-body Kepler problem (see Eq.~\ref{e:two-body-newton-ode}). The same applies to $m_2$ and $m_3$. So if $z_a(0)$ denote the initial positions, the curves $z_a(t) = z(t) z_a(0)$ are solutions of Newton's equations for three bodies provided $z(t)$ is a Keplerian orbit for an appropriate two-body problem. In other words, each mass traverses a rescaled Keplerian orbit about the common centre of mass. A similar analysis applies to the Euler collinear solutions as well: locations of the masses is determined by the requirement that the force on each one is toward the CM and proportional to the distance from it (see Box 5 on central configurations).

%------------------

\begin{center}
	\begin{mdframed}
{\bf Box 5: Central configurations:} Three-body configurations in which the acceleration of each particle points towards the CM and is proportional to its distance from the CM (${\bf a}_b= \om^2 (\bfR_{\rm CM} - \bfr_b)$ for $b = 1,2,3$) are called `central configurations'. A central configuration rotating at angular speed $\om$ about the CM automatically satisfies the equations of motion (\ref{e:newtonian-3body-ODE}). Euler collinear and Lagrange equilateral configurations are the only central configurations in the three-body problem. In 1912, Karl Sundmann showed that triple collisions are asymptotically central configurations. 
%The same also holds for $n$ bodies.
%They play a role in finding explicit solutions, studying near collision trajectories and topology of the integral manifolds.
	\end{mdframed}
\end{center}

%----------------

%In general, for both Euler and Lagrange solutions:
%	\beq
%	F_i = -\grad_i V = -\la m_i \left(\bfr_i - \bfr_{\rm CM}\right) \quad \text{where} \quad \bfr_{\rm CM}= \fr{\sum m_i \bfr_i}{\sum m_i}, \la = \fr{-V}{\sum m_i |\bfr_i - \bfr_{\rm CM}|^2}.
%	\eeq
%For the first body in Lagrange solution, $V = \al_1/ |x_1 - x_{\rm CM}|,$ $\sum m_i |x_i - x_{\rm CM}|^2 = \beta_1 |x_1 - x_{\rm CM}|^2$:
%	\beq
%	F_1 = \fr{V}{\sum m_i |x_i - x_{\rm CM}|^2} m_i \left(x_1 - x_{\rm CM}\right) = \fr{\al_1}{\beta_1} \fr{m_i \left(x_1 - x_{\rm CM}\right)}{|x_1 - x_{\rm CM}|^3}.
%	\eeq
%Here $\al_i$ and $\beta_i$ depend on masses of the bodies and so are independent of time. So solutions of the sort $x_i(t) = x(t) x_i(0)$ solve the Newton's laws of motion for three bodies where $x(t)$ is a keplerian solution for some corresponding two-body problem.

%---------------
\section{Restricted three-body problem}
%---------------

\begin{figure}[h] 
	\center
	\includegraphics[width=5cm]{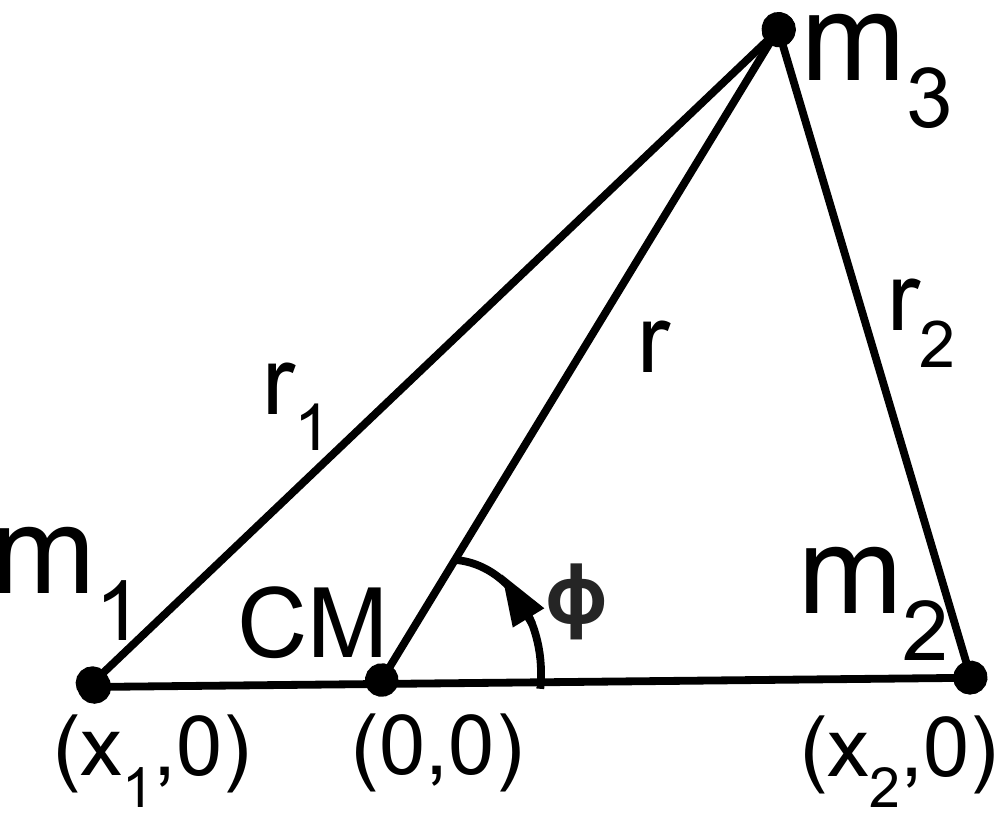}
	\caption{\footnotesize The secondary $m_3$ in the co-rotating frame of primaries $m_1$ and $m_2$ in the restricted three-body problem. The origin is located at the center of mass of $m_1$ and $m_2$ which coincides with the CM of the system since $m_3 \ll m_{1,2}$.}
	\label{f:restricted-3body-setup}
\end{figure}

The restricted three-body problem is a simplified version of the three-body problem where one of the masses $m_3$ is assumed much smaller than the primaries $m_1$ and $m_2$. Thus, $m_1$ and $m_2$ move in Keplerian orbits which are not affected by $m_3$. The Sun-Earth-Moon system provides an example where we further have $m_2 \ll m_1$. In the planar circular restricted three-body problem, the primaries move in fixed circular orbits around their common CM with angular speed $\Omega = (G (m_1 + m_2)/d^3 )^{1/2}$ given by Kepler's third law and $m_3$ moves in the same plane as $m_1$ and $m_2$. Here, $d$ is the separation between primaries. This system has $2$ degrees of freedom associated to the planar motion of $m_3$, and therefore a 4-dimensional phase space just like the planar Kepler problem for the reduced mass. However, unlike the latter which has three conserved quantities (energy, $z$-component of angular momentum and direction of LRL vector) and is exactly solvable, the planar restricted three-body problem has only one known conserved quantity, the `Jacobi integral', which is the energy of $m_3$ in the co-rotating (non-inertial) frame of the primaries:
	\beq 
	 E = \left[ \half m_3 \dot r^2 + \half m_3 r^2 \dot \phi^2 \right] - \half m_3 \Om^2 r^2 - G m_3 \left( \fr{m_1}{r_1} + \fr{m_2}{r_2} \right) \equiv T + V_{\rm eff}.
	\eeq
Here, $(r,\phi)$ are the plane polar coordinates of $m_3$ in the co-rotating frame of the primaries with origin located at their center of mass while $r_1$ and $r_2$ are the distances of $m_3$ from $m_1$ and $m_2$ (see Fig.~\ref{f:restricted-3body-setup}). The `Roche' effective potential $V_{\rm eff}$, named after the French astronomer \'Edouard Albert Roche, is a sum of centrifugal and gravitational energies due to $m_1$ and $m_2$.

A system with $n$ degrees of freedom needs at least $n$ constants of motion to be exactly solvable\footnote{A Hamiltonian system with $n$ degrees of freedom is exactly solvable in the sense of Liouville if it possesses $n$ independent conserved quantities in involution, i.e., with vanishing pairwise Poisson brackets (see Boxes 6 and 10).}. For the restricted 3-body problem, Henri Poincar\'e (1889) proved the nonexistence of any conserved quantity (other than $E$) that is analytic in small mass ratios ($m_3/m_2$ and $(m_3+m_2)/m_1$) and orbital elements ($\bfJ_1$, $M_1 \dot \bfJ_1$, $\bfJ_2$ and $M_2 \dot \bfJ_2$)  \cite{diacu-holmes,musielak-quarles,barrow-green-poincare-three-body}. This was an extension of a result of Heinrich Bruns who had proved in 1887 the nonexistence of any new conserved quantity algebraic in Cartesian coordinates and momenta for the general three-body problem \cite{whittaker}. Thus, roughly speaking, Poincar\'e showed that the restricted three-body problem is not exactly solvable. In fact, as we outline in \S\ref{s:delaunay-hill-poincare}, he discovered that it displays chaotic behavior.

\begin{figure}[h] 
	\center
	\includegraphics[width=5cm]{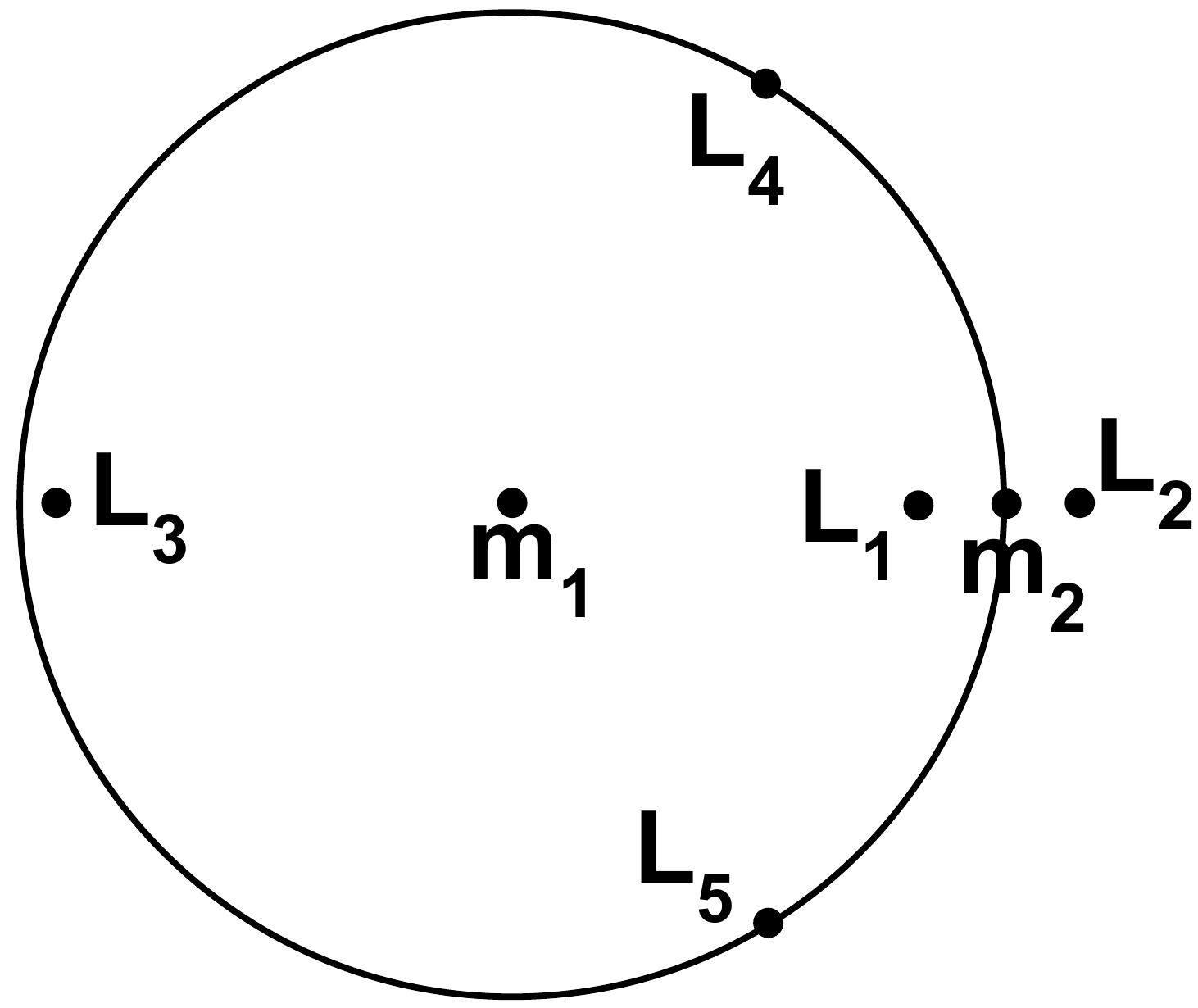}
	\caption{\footnotesize The positions of Euler $(L_{1,2,3})$ and Lagrange $(L_{4,5})$ points when $m_1 \gg m_2 \gg m_3$. $m_2$ is in an approximately circular orbit around $m_1$. $L_3$ is almost diametrically opposite to $m_2$ and a bit closer to $m_1$ than $m_2$ is. $L_1$ and $L_2$ are symmetrically located on either side of $m_2$. $L_4$ and $L_5$ are equidistant from $m_1$ and $m_2$ and lie on the circular orbit of $m_2$.}
	\label{f:euler-lagrange-points}
\end{figure}

\begin{figure}[h] 
	\center
	\includegraphics[width=8cm]{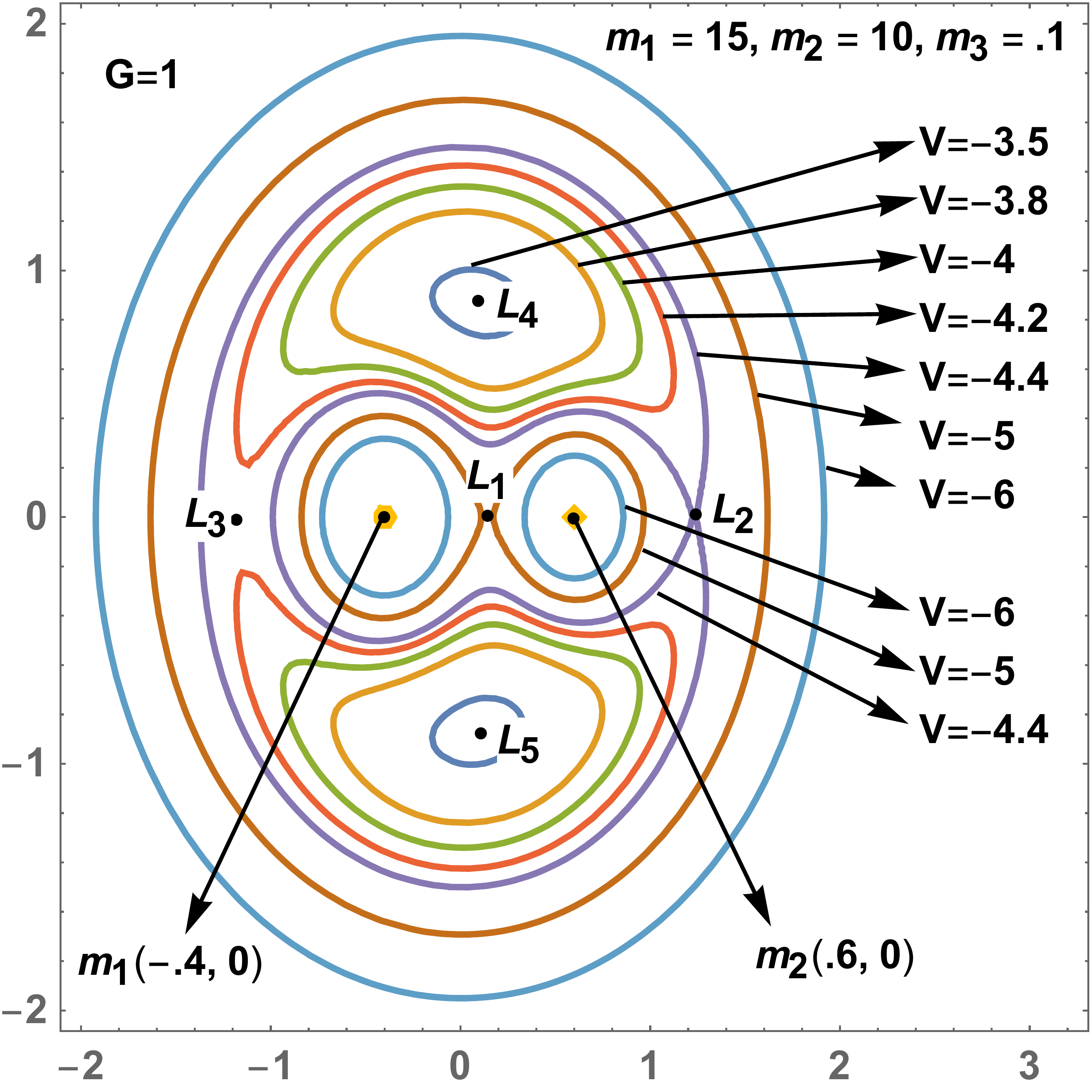}
	\caption{\footnotesize Level curves of the Roche effective potential energy $V_{\rm eff}$ of $m_3$ in the co-rotating frame of the primaries $m_1$ and $m_2$ in the circular restricted three-body problem for $G = 1$, $m_1 = 15, m_2 = 10$ and $m_3 = .1$. Lagrange points $L_{1-5}$ are at extrema of $V_{\rm eff}$. The trajectory of $m_3$ for a given energy $E$ must lie in the Hill region defined by $V_{\rm eff}(x,y) \leq E$. E.g., for $E=-6$, the Hill region is the union of two neighborhoods of the primaries and a neighborhood of the point at infinity. The lobes of the $\infty$-shaped level curve passing through $L_1$ are called Roche's lobes. The saddle point $L_1$ is like a mountain pass through which material could pass between the lobes.}
	\label{f:effective-potential}
\end{figure}

%---------------
{\noindent \bf Euler and Lagrange points}\footnote{Lagrange points $L_{1-5}$ are also called libration (literally, balance) points.} (denoted $L_{1-5}$) of the restricted three-body problem are the locations of a third mass ($m_3 \ll m_1, m_2$) in the co-rotating frame of the primaries $m_1$ and $m_2$ in the Euler and Lagrange solutions (see Fig.~\ref{f:euler-lagrange-points}). Their stability would allow an asteroid or satellite to occupy a Lagrange point. Euler points $L_{1,2,3}$ are saddle points of the Roche potential while $L_{4,5}$ are maxima (see Fig.~\ref{f:effective-potential}). This suggests that they are all unstable. However, $V_{\rm eff}$ does not include the effect of the Coriolis force since it does no work. A more careful analysis shows that the Coriolis force stabilizes $L_{4,5}$. It is a bit like a magnetic force which does no work but can stabilize a particle in a Penning trap. Euler points are always unstable\footnote{Stable `Halo' orbits around Euler points have been found numerically.} while the Lagrange points $L_{4,5}$ are stable to small perturbations iff $(m_1+m_2)^2 \geq 27 m_1 m_2$ \cite{symon}. More generally, in the unrestricted three-body problem, the Lagrange equilateral solutions are stable iff
	\beq
	(m_1 + m_2 + m_3)^2 \geq 27(m_1 m_2 + m_2 m_3 + m_3 m_1).	
	\eeq
The above criterion due to Edward Routh (1877) is satisfied if one of the masses dominates the other two. For instance, $L_{4,5}$ for the Sun-Jupiter system are stable and occupied by the Trojan asteroids.

%---------------

%-------------------
\section{Planar Euler three-body problem}
%-------------------

\begin{figure}[h] 
\center
\includegraphics[width=6cm]{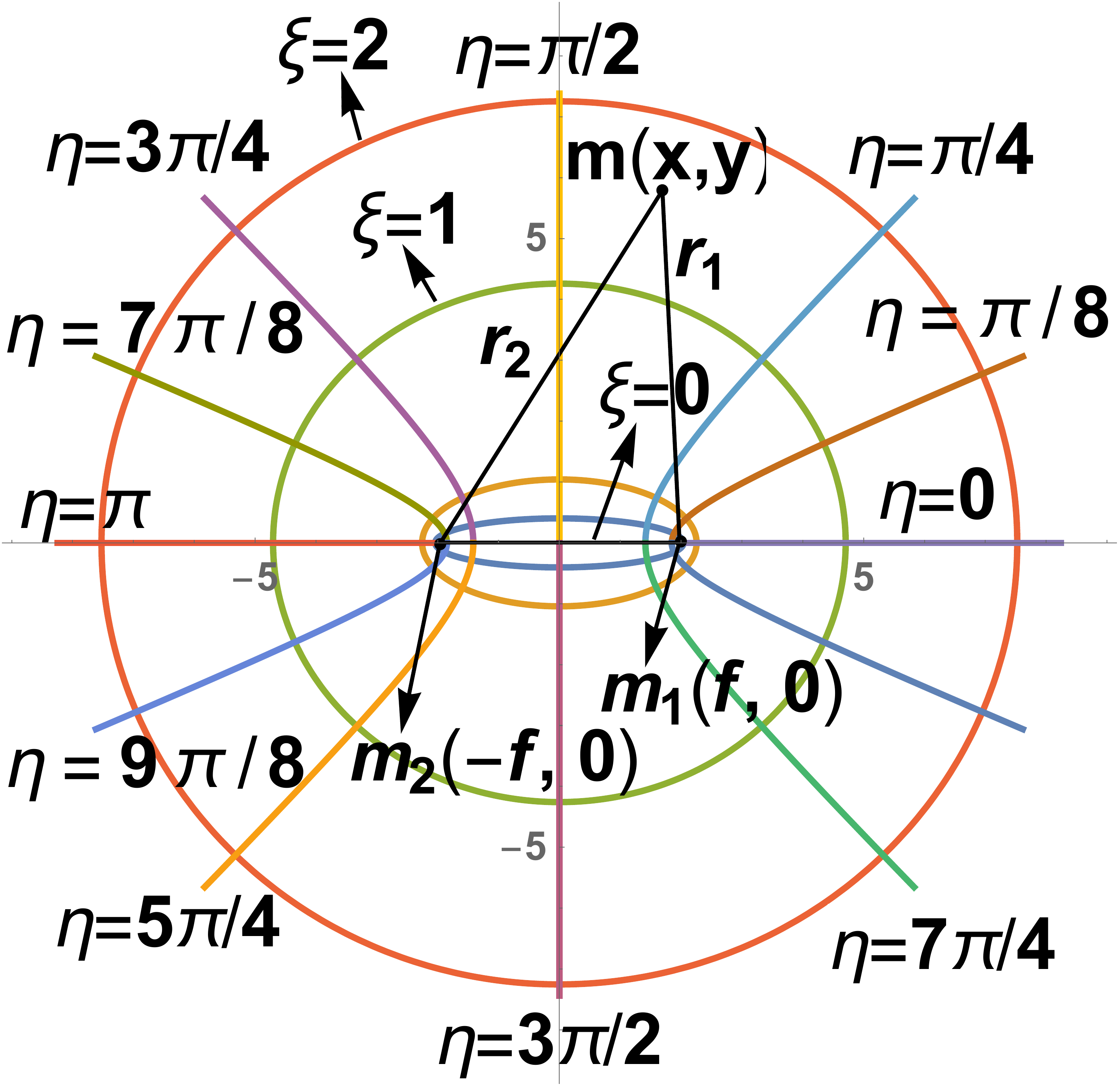}
\caption{\footnotesize Elliptical coordinate system for the Euler 3-body problem. Two masses are at the foci $(\pm f,0)$ of an elliptical coordinate system with $f=2$ on the $x$-$y$ plane. The level curves of $\xi$ and $\eta$ (confocal ellipses and hyperbolae) are indicated. }
\label{f:elliptic-coordinates}
\end{figure}

Given the complexity of the restricted three-body problem, Euler (1760) proposed the even simpler problem of a mass $m$ moving in the gravitational potential of two {\it fixed} masses $m_1$ and $m_2$. Initial conditions can be chosen so that $m$ always moves on a fixed plane containing $m_1$ and $m_2$. Thus, we arrive at a one-body problem with two degrees of freedom and energy
	\beq
	E = \half m \left(\dot x^2 + \dot y^2 \right) -\frac{\mu_1}{r_1} - \frac{\mu_2}{r_2}.
	\label{e:euler-three-body-energy}
	\eeq
Here, $(x,y)$ are the Cartesian coordinates of $m$, $r_a$ the distances of $m$ from $m_a$ and $\mu_a = G m_a m$ for $a = 1,2$ (see Fig.~\ref{f:elliptic-coordinates}). Unlike in the restricted three-body problem, here the rest-frame of the primaries is an inertial frame, so there are no centrifugal or Coriolis forces. This simplification allows the Euler three-body problem to be exactly solved.

Just as the Kepler problem simplifies in plane-polar coordinates $(r, \tht)$ centered at the CM, the Euler 3-body problem simplifies in an elliptical coordinate system $(\xi, \eta)$. The level curves of $\xi$ and $\eta$ are mutually orthogonal confocal ellipses and hyperbolae (see Fig.~\ref{f:elliptic-coordinates}) with the two fixed masses at the foci $2f$ apart:
	\beq
	x = f \: \cosh\xi \: \cos\eta \quad \text{and} \quad 
	y = f \: \sinh\xi \: \sin\eta.
	\label{e:elliptical-coordinates-transformation}
	\eeq
Here, $\xi$ and $\eta$ are like the radial distance $r$ and angle $\tht$, whose level curves are mutually orthogonal concentric circles and radial rays. The distances of $m$ from $m_{1,2}$ are $r_{1,2}= f (\cosh \xi \mp \cos \eta)$.

The above confocal ellipses and hyperbolae are Keplerian orbits when a single fixed mass ($m_1$ or $m_2$) is present at one of the foci $(\pm f,0)$. Remarkably, these Keplerian orbits survive as orbits of the Euler 3-body problem. This is a consequence of Bonnet's theorem, which states that if a curve is a trajectory in two separate force fields, it remains a trajectory in the presence of both. If $v_1$ and $v_2$ are the speeds of the Keplerian trajectories when only $m_1$ or $m_2$ was present, then $v = \sqrt{v_1^2 + v_2^2}$ is the speed when both are present. 

Bonnet's theorem however does not give us all the trajectories of the Euler 3-body problem. More generally, we may integrate the equations of motion by the method of separation of variables in the Hamilton-Jacobi equation (see \cite{mukunda-hamilton} and Boxes 6, 7 \& 8). The system possesses {\it two} independent conserved quantities: energy and Whittaker's constant \footnote{When the primaries coalesce at the origin  ($f \to 0$), Whittaker's constant reduces to the conserved quantity $\bfL^2$ of the planar 2-body problem.} \cite{gutzwiller-book, whittaker}
	\beq
	w = \bfL_1 \cdot \bfL_2 + 2 m f \left( -\mu_1\cos\theta_1 + \mu_2\cos\theta_2 \right) =  m^2 r_1^2 \, r_2^2 \; \dot \tht_1 \dot \tht_2 + 2f m  \left( -\mu_1\cos\theta_1 + \mu_2\cos\theta_2 \right).
	\label{e:whittakers-constant}
	\eeq
Here, $\tht_a$ are the angles between the position vectors $\bfr_a$ and the positive $x$-axis and $\bfL_{1,2} = m r_{1,2}^2 \dot \tht_{1,2} \hat z$ are the angular momenta about the two force centers (Fig.~\ref{f:elliptic-coordinates}). Since $w$ is conserved, it Poisson commutes with the Hamiltonian $H$. Thus, the planar Euler 3-body problem has two degrees of freedom and two conserved quantities in involution. Consequently, the system is integrable in the sense of Liouville.

More generally, in the three-dimensional Euler three-body problem, the mass $m$ can revolve (non-uniformly) about the line joining the force centers ($x$-axis) so that its motion is no longer confined to a plane. Nevertheless, the problem is exactly solvable as the equations admit three independent constants of motion in involution: energy, Whittaker's constant and the $x$ component of angular momentum \cite{gutzwiller-book}.

\begin{center}
	\begin{mdframed}
{\bf Box 6: Canonical transformations:} We have seen that the  Kepler problem is more easily solved in polar coordinates and momenta $(r, \tht, p_r, p_\tht)$ than in Cartesian phase space variables $(x, y, p_x, p_y)$. This change is an example of a canonical transformation (CT). More generally, a CT is a change of canonical phase space variables $(\bfq, \bfp) \to ({\bf Q} (\bfp, \bfq, t), {\bf P}(\bfp, \bfq, t))$ that preserves the form of Hamilton's equations. For one degree of freedom, Hamilton's equations $\dot q = \dd{H}{p}$ and $\dot p = -\dd{H}{q}$ become $\dot Q = \dd{K}{P}$ and $\dot P = -\dd{K}{Q}$ where $K(Q,P,t)$ is the new Hamiltonian (for a time independent CT, the old and new Hamiltonians are related by substitution: $H(q,p) = K(Q(q,p),P(q,p))$). The form of Hamilton's equations is preserved provided the basic Poisson brackets do not change i.e.,
	\beq
	\{ q,p \} = 1, \;\; \{ q,q \} = \{ p,p\} = 0 \quad \imply \quad \{ Q,P \} = 1, \;\; \{ Q,Q \} = \{ P,P \} = 0.
	\eeq
Here, the Poisson bracket of two functions on phase space $f(q,p)$ and $g(q,p)$ is defined as
	\beq
	\{ f(q,p), g(q,p) \} = \dd{f}{q} \dd{g}{p} - \dd{f}{p} \dd{g}{q}.
	\eeq
For one degree of freedom, a CT is simply an area and orientation preserving transformation of the $q$-$p$ phase plane. Indeed, the condition $\{ Q, P \} = 1$ simply states that the Jacobian determinant $J = \det \left( \dd{Q}{q}, \dd{Q}{p} \;\vert\; \dd{P}{q}, \dd{P}{p} \right) = 1$ so that the new area element $dQ \, dP = J \, dq \, dp$ is equal to the old one. A CT can be obtained from a suitable generating function, say of the form $S(q,P,t)$, in the sense that the equations of transformation are given by partial derivatives of $S$:
	\beq
	p = \dd{S}{q}, \quad Q = \dd{S}{P} \quad \text{and} \quad K = H + \dd{S}{t}.
	\eeq
For example, $S = qP$ generates the identity transformation ($Q = q$ and $P = p$) while $S = - qP$ generates a rotation of the phase plane by $\pi$ ($Q = -q$  and $P = -p$).
	\end{mdframed}
\end{center}

%-------------------

\begin{center}
	\begin{mdframed}
{\bf Box 7: Hamilton Jacobi equation:} The Hamilton Jacobi (HJ) equation is an alternative formulation of Newtonian dynamics. Let $i = 1, \ldots, n$ label the degrees of freedom of a mechanical system. Cyclic coordinates $q^i$ (i.e., those that do not appear in the Hamiltonian $H(\bfq,\bfp,t)$ so that $\partial H/ \pdr q^i = 0$) help to understand Newtonian trajectories, since their conjugate momenta $p_i$ are conserved ($\dot p_i = \dd{H}{q^i} = 0$). If all coordinates are cyclic, then each of them evolves linearly in time: $q^i(t) = q^i(0) + \dd{H}{p_i} t$. Now time-evolution is {\it even simpler} if $\dd{H}{p_i} = 0$ for all $i$ as well, i.e., if $H$ is independent of both coordinates and momenta! In the HJ approach, we find a CT from old phase space variables $(\bfq, \bfp)$ to such a coordinate system $(\bfQ,\bfP)$ in which the new Hamiltonian $K$ is a constant (which can be taken to vanish by shifting the zero of energy). The HJ equation is a nonlinear, first-order partial differential equation for Hamilton's principal function $S(\bfq,\bfP,t)$ which generates the canonical transformation from $(\bfq,\bfp)$ to $(\bfQ,\bfP)$. As explained in Box 6, this means $p_i = \dd{S}{q^i}$, $Q^j = \dd{S}{P_j}$ and $K = H + \dd{S}{t}$. Thus, the HJ equation 	\beq
	H\left(\bfq, \dd{S}{\bfq},t \right) + \dd{S}{t} = 0
	\eeq
is simply the condition for the new Hamiltonian $K$ to vanish. If $H$ is time-independent, we may `separate' the time-dependence of $S$ by writing $S(\bfq,\bfP,t) = W(\bfq,\bfP) - Et$ where the `separation constant' $E$ may be interpreted as energy. Thus, the time independent HJ-equation for Hamilton's characteristic function $W$ is
	\beq
	H\left(\bfq,\frac{\partial W}{\partial \bfq}\right) = E.
	\label{e:time-indep-HJ}
	\eeq
E.g., for a particle in a potential $V(\bfq)$, it is the equation $\ov{2m}\left( \fr{\pdr W}{\pdr \bfq}\right)^2 + V(\bfq) = E$. By solving (\ref{e:time-indep-HJ}) for $W$, we find the desired canonical transformation to the new conserved coordinates $\bfQ$ and momenta $\bfP$. By inverting the relation $(q,p) \mapsto (Q,P)$ we find $(q^i(t),p_j(t))$ given their initial values. $W$ is said to be a {\it complete integral} of the HJ equation if it depends on $n$ constants of integration, which may be taken to be the new momenta $P_1, \ldots, P_n$. When this is the case, the system is said to be integrable via the HJ equation. However, it is seldom possible to find such a complete integral. In favorable cases, {\it separation of variables} can help to solve the HJ equation (see Box 8).
	\end{mdframed}
\end{center}

%---------------

\begin{center}
	\begin{mdframed}
{\bf Box 8:} {\bf Separation of variables:} In the planar Euler 3-body problem, Hamilton's characteristic function $W$ depends on the two `old' elliptical coordinates $\xi$ and $\eta$. The virtue of elliptical coordinates is that the time-independent HJ equation can be solved by separating the dependence of $W$ on $\xi$ and $\eta$: $W(\xi, \eta) = W_1(\xi) + W_2(\eta)$. Writing the energy (\ref{e:euler-three-body-energy}) in elliptical coordinates (\ref{e:elliptical-coordinates-transformation})  and using $p_\xi = W_1'(\xi)$ and $p_\eta = W_2'(\eta)$, the time-independent HJ equation (\ref{e:time-indep-HJ}) becomes 
	\beq
	E = \frac{W_1'(\xi)^2 + W_2'(\eta)^2 - 2mf(\mu_1+\mu_2)\cosh\xi -2mf(\mu_1-\mu_2)\cos\eta}{2mf^2(\cosh^2\xi-\cos^2\eta)}.
	\eeq
Rearranging,
	\beq
	W_1'^2 - 2Emf^2\cosh^2\xi - 2mf(\mu_1+\mu_2)\cosh\xi = -W_2'^2  -2Emf^2\cos^2\eta + 2mf(\mu_1-\mu_2)\cos\eta.
	\eeq
Since the LHS and RHS are functions only of $\xi$ and $\eta$ respectively, they must both be equal to a `separation constant' $\al$. Thus, the HJ partial differential equation separates into a pair of decoupled ODEs for $W_1(\xi)$ and $W_2(\eta)$. The latter may be integrated using elliptic functions. Note that Whittaker's constant $w$ (\ref{e:whittakers-constant}) may be expressed as $w = - 2 m f^2 E - \al$.
	\end{mdframed}
\end{center}

%\pt Pierre-Simon Laplace (1785) accounted for the effect of resonances in Sun-Jupiter-Saturn system ($5 T_{J} \approx 2 T_{S}$) to understand deviations from Keplerian orbits. He prepared a grand synthesis by deriving all solar system quantities observed at the time from Newton's laws.

%--------------
\section{Some landmarks in the history of the 3-body problem} 
\label{s:delaunay-hill-poincare}
%--------------

The importance of the three-body problem lies in part in the developments that arose from attempts to solve it \cite{diacu-holmes,musielak-quarles}. These have had an impact all over astronomy, physics and mathematics. 

Can planets collide, be ejected from the solar system or suffer significant deviations from their Keplerian orbits? This is the question of the stability of the solar system. In the $18^{\rm th}$ century, Pierre-Simon Laplace and J. L. Lagrange obtained the first significant results on stability. They showed that to first order in the ratio of planetary to solar masses ($M_p/M_S$), there is no unbounded variation in the semi-major axes of the orbits, indicating stability of the solar system. Sim\'eon Denis Poisson extended this result to second order in $M_p/M_S$. However, in what came as a surprise, the Romanian Spiru Haretu (1878) overcame significant technical challenges to find secular terms (growing linearly and quadratically in time) in the semi-major axes at third order! This was an example of a perturbative expansion, where one expands a physical quantity in powers of a small parameter (here the semi-major axis was expanded in powers of $M_p/M_S \ll 1$). Haretu's result however did not prove instability as the effects of his secular terms could cancel out (see Box 9 for a simple example). But it effectively put an end to the hope of proving the stability/instability of the solar system using such a perturbative approach.

The development of Hamilton's mechanics and its refinement in the hands of Carl Jacobi was still fresh when the French dynamical astronomer Charles Delaunay (1846) began the first extensive use of canonical transformations (see Box 6) in perturbation theory \cite{gutzwiller-three-body}. The scale of his hand calculations is staggering: he applied a succession of 505 canonical transformations to a $7^{\rm th}$ order perturbative treatment of the three-dimensional elliptical restricted three-body problem. He arrived at the equation of motion for $m_3$ in Hamiltonian form using $3$ pairs of canonically conjugate orbital variables (3 angular momentum components, the true anomaly, longitude of the ascending node and distance of the ascending node from perigee). He obtained the latitude and longitude of the moon in trigonometric series of about $450$ terms with secular terms (see Box 9) eliminated. It wasn't till 1970-71 that Delaunay's heroic calculations were checked and extended using computers at the Boeing Scientific Laboratories \cite{gutzwiller-three-body}!

The Swede Anders Lindstedt (1883) developed a systematic method to approximate solutions to nonlinear ODEs when naive perturbation series fail due to secular terms (see Box 9). The technique was further developed by Poincar\'e. Lindstedt assumed the series to be generally convergent, but Poincar\'e soon showed that they are divergent in most cases. Remarkably, nearly 70 years later, Kolmogorov, Arnold and Moser showed that in many of the cases where Poincar\'e's arguments were inconclusive, the series are in fact convergent, leading to the celebrated KAM theory of integrable systems subject to small perturbations (see Box 10).

%-----------------------------

\begin{center}
	\begin{mdframed}
{\bf Box 9: Poincar\'e-Lindstedt method:} The Poincar\'e-Lindstedt method is an approach to finding series solutions to a system such as the anharmonic oscillator $\ddot x + x + g x^3 = 0$, which for small $g$, is a perturbation of the harmonic oscillator $m \ddot x + k x = 0$ with mass $m = 1$ and spring constant $k = 1$. The latter admits the periodic solution $x_0(t) = \cos t$ with initial conditions $x(0) = 1$, $\dot x(0) = 0$. For a small perturbation $0 < g \ll 1$, expanding $x(t) = x_0(t) + g x_1(t) + \cdots$ in powers of $g$ leads to a linearized equation for $x_1(t)$
	\beq 
	\label{e:x1-lindstedt}
	\ddot x_1 + x_1 + \cos^3 t = 0.
	\eeq
However, the perturbative solution
	\beq
	x(t) = x_0 + g x_1 + {\cal O}(g^2) = \cos t + g \left[ \ov{32} (\cos 3t - \cos t) - \fr{3}{8} t \sin t \right] + {\cal O}(g^2)
	\eeq
is unbounded due to the linearly growing {\it secular} term $(-3/8)t \sin t$. This is unacceptable as the energy $E =\half \dot x^2 + \half x^2 + \ov{4} g x^4$ must be conserved and the particle must oscillate between turning points of the potential $V = \half x^2 + \fr{g}{4} x^4$. The Poincar\'e-Lindstedt method avoids this problem by looking for a series solution of the form
	\beq
	x(t) = x_0(\tau) + g \tl x_1(\tau) + \cdots
	\eeq
where $\tau = \om t$ with $\om = 1 + g \om_1 + \cdots$. The constants $\om_1, \om_2, \cdots$ are chosen to ensure that the coefficients of the secular terms at order $g, g^2, \cdots$ vanish. In the case at hand we have 
	\beq
	x(t) = \cos (t + g \om_1 t) + g \tl x_1(t) + {\cal O}(g^2)
	= \cos t + g \tl {\tl x}_1(t) + {\cal O}(g^2) \quad \text{where} \quad \tl {\tilde x}_1(t) = \tl x_1(t) - \om_1 t \sin t.
	\eeq
$\tl {\tilde x}_1$ satisfies the same equation (\ref{e:x1-lindstedt}) as $x_1$ did, leading to 
	\beq
	\tl x_1(t) = \ov{32} (\cos 3t - \cos t) + \left(\om_1 - \fr{3}{8} \right) t \sin t.
	\eeq
The choice $\om_1 = 3/8$ ensures cancellation of the secular term at order $g$, leading to the approximate bounded solution
	\beq
	x(t) = \cos \left(t + \frac{3}{8} g t \right) + \frac{g}{32} \left(\cos 3t - \cos t \right) + {\cal O}\left(g^2 \right).
	\eeq
	\end{mdframed}
\end{center}

%---------------

%-----------------------------
\begin{center}
	\begin{mdframed}
{\bf Box 10: Action-angle variables and invariant tori:} Time evolution is particularly simple if all the generalized coordinates $\tht^j$ are cyclic so that their conjugate momenta $I_j$ are conserved: $\dot I_j = - \dd{H}{\tht^j} = 0$. A Hamiltonian system with $n$ degrees of freedom is integrable in the sense of Liouville if it admits $n$ canonically conjugate ($\{ \tht^j, I_k \} = \del^j_k$\footnote{The Kronecker symbol $\del^j_k$ is equal to one for $j = k$ and zero otherwise}) pairs of phase space variables $(\tht^j, I_j)$ with all the $\tht^j$ cyclic, so that its Hamiltonian depends only on the momenta, $H = H({\bf I})$. Then the `angle' variables $\tht^j$ evolve linearly in time $(\tht^j(t) = \tht^j(0) + \om^j \: t)$ while the momentum or `action' variables $I_j$ are conserved. Here, $\om^j = \dot \tht^j = \dd{H}{I_j}$ are $n$ constant frequencies. Typically the angle variables are periodic, so that the $\tht^j$ parametrize circles. The common level sets of the action variables $I_j = c_j$ are therefore a family of tori that foliate the phase space. Recall that a torus is a Cartesian product of circles. For instance, for one degree of freedom, $\tht^1$ labels points on a circle $S^1$ while for 2 degrees of freedom, $\tht^1$ and $\tht^2$ label points on a 2-torus $S^1 \times S^1$ which looks like a vada or doughnut. Trajectories remain on a fixed torus determined by the initial conditions. Under a sufficiently small and smooth perturbation $H({\bf I}) + g H'({\bf I}, {\vec \tht})$, Andrei Kolmogorov, Vladimir Arnold and J\"urgen Moser showed that some of these `invariant' tori survive provided the frequencies $\om^i$ are sufficiently `non-resonant' or `incommensurate' (i.e., their integral linear combinations do not get `too small').
	\end{mdframed}
\end{center}

%-----------------------------

George William Hill was motivated by discrepancies in lunar perigee calculations. His celebrated paper on this topic was published in 1877 while working with Simon Newcomb at the American Ephemeris and Nautical Almanac\footnote{Simon Newcomb's project of revising all the orbital data in the solar system established the missing $42''$ in the $566''$ centennial precession of Mercury's perihelion. This played an important role in validating Einstein's general theory of relativity.}. He found a new family of periodic orbits in the circular restricted (Sun-Earth-Moon) 3-body problem by using a frame rotating with the Sun's angular velocity instead of that of the Moon. The solar perturbation to lunar motion around the Earth results in differential equations with periodic coefficients. He used Fourier series to convert these ODEs to an infinite system of linear algebraic equations and developed a theory of infinite determinants to solve them and obtain a rapidly converging series solution for lunar motion. He also discovered new `tight binary' solutions to the 3-body problem where two nearby masses are in nearly circular orbits around their center of mass CM$_{12}$, while CM$_{12}$ and the far away third mass in turn orbit each other in nearly circular trajectories.

\begin{figure}	
	\centering
	\begin{subfigure}[t]{3cm}
		\centering
		\includegraphics[width=3cm]{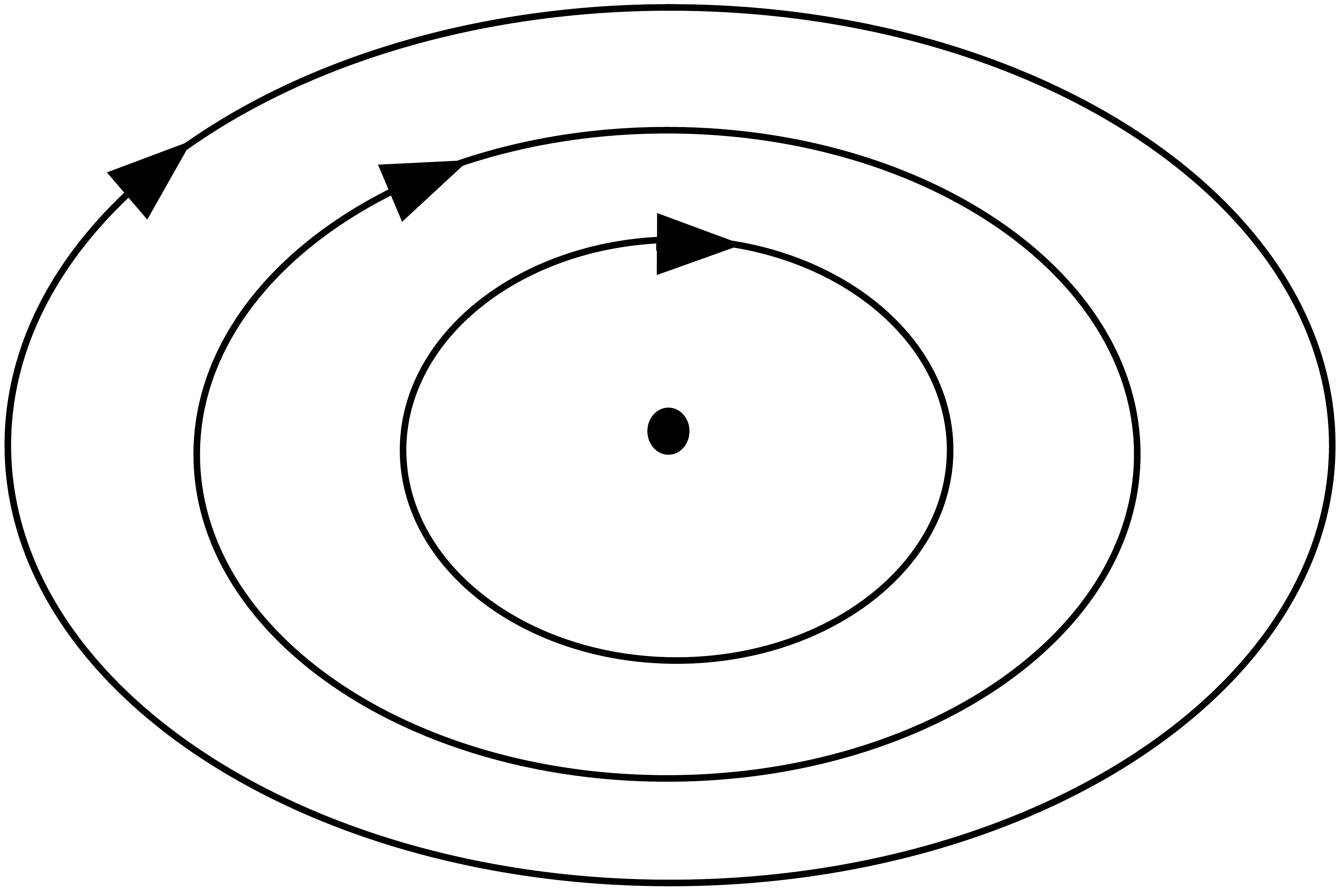}
		\caption{\footnotesize center}
		\label{f:centre}		
	\end{subfigure}
	\quad
	\begin{subfigure}[t]{3cm}
		\centering
		\includegraphics[width=3cm]{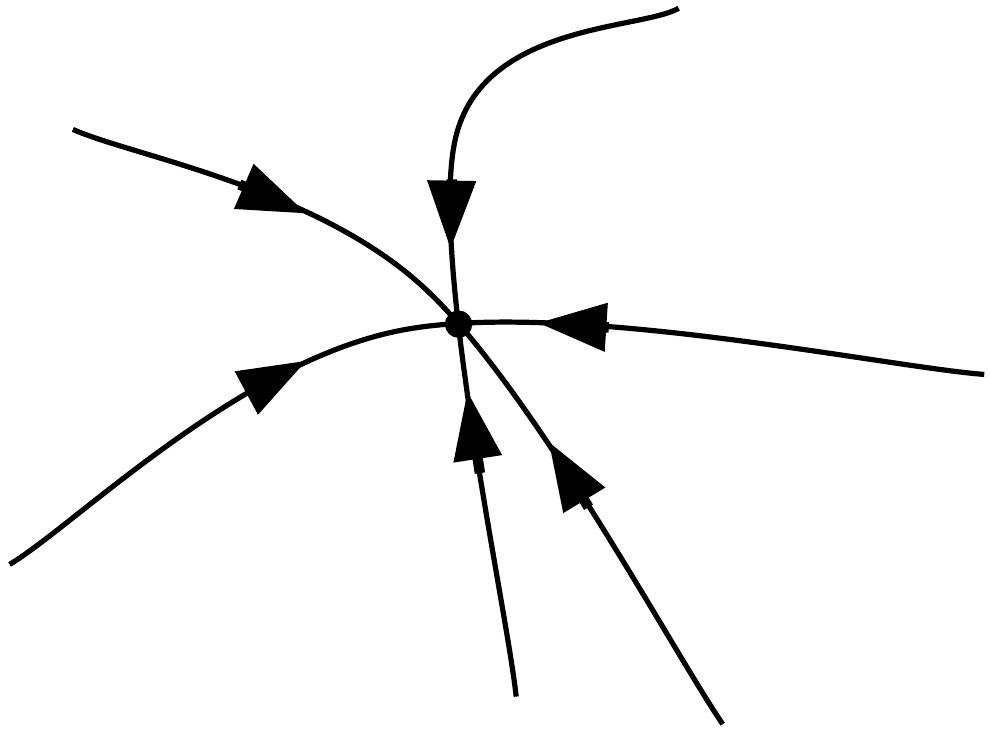}
		\caption{\footnotesize (stable) node}
		\label{f:node}		
	\end{subfigure}		
	\begin{subfigure}[t]{3cm}
		\centering
		\includegraphics[width=2.1cm]{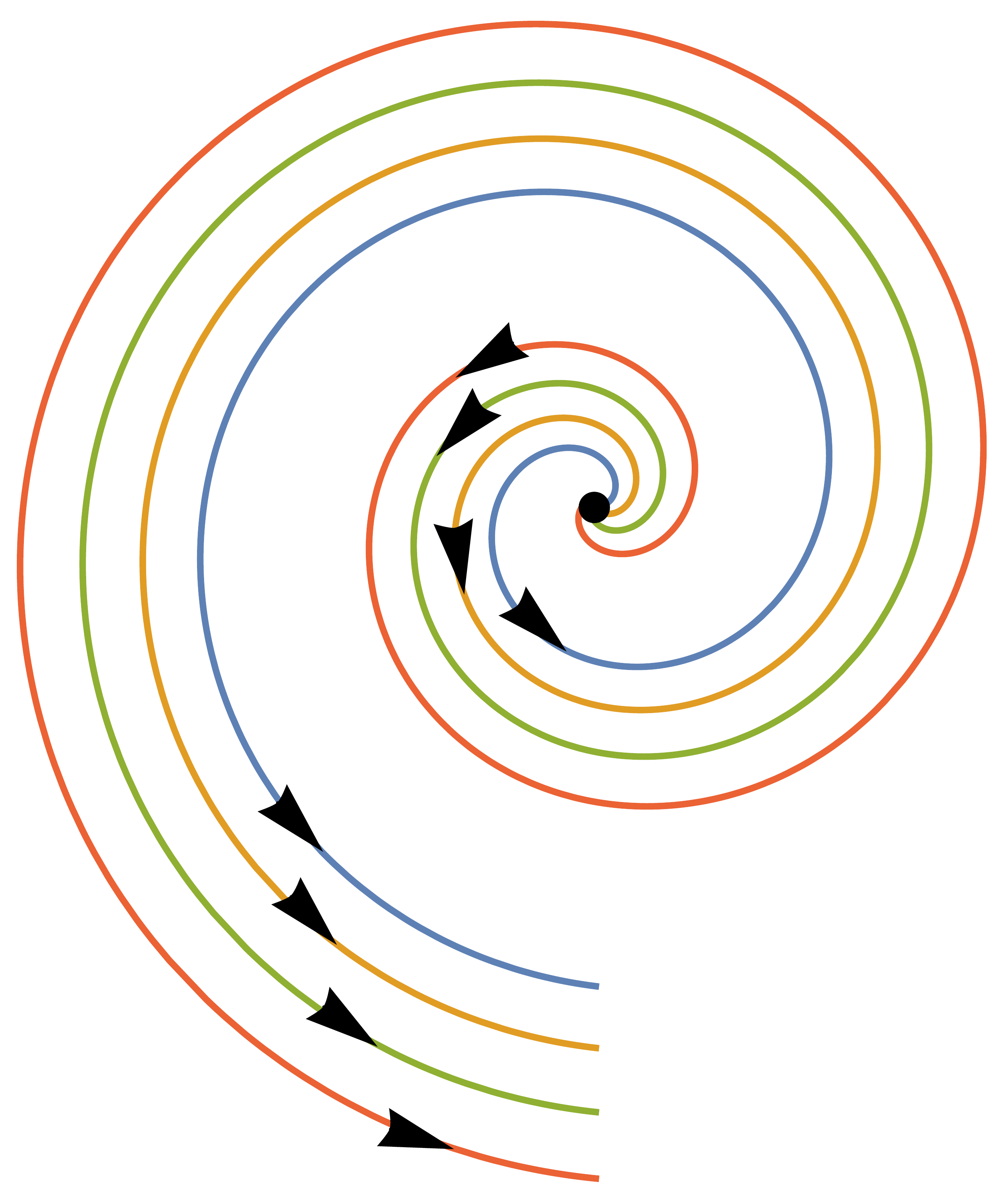}
		\caption{\footnotesize (unstable) focus}
		\label{f:focus}		
	\end{subfigure}
	\quad	
	\begin{subfigure}[t]{3cm}
		\centering
		\includegraphics[width=3cm]{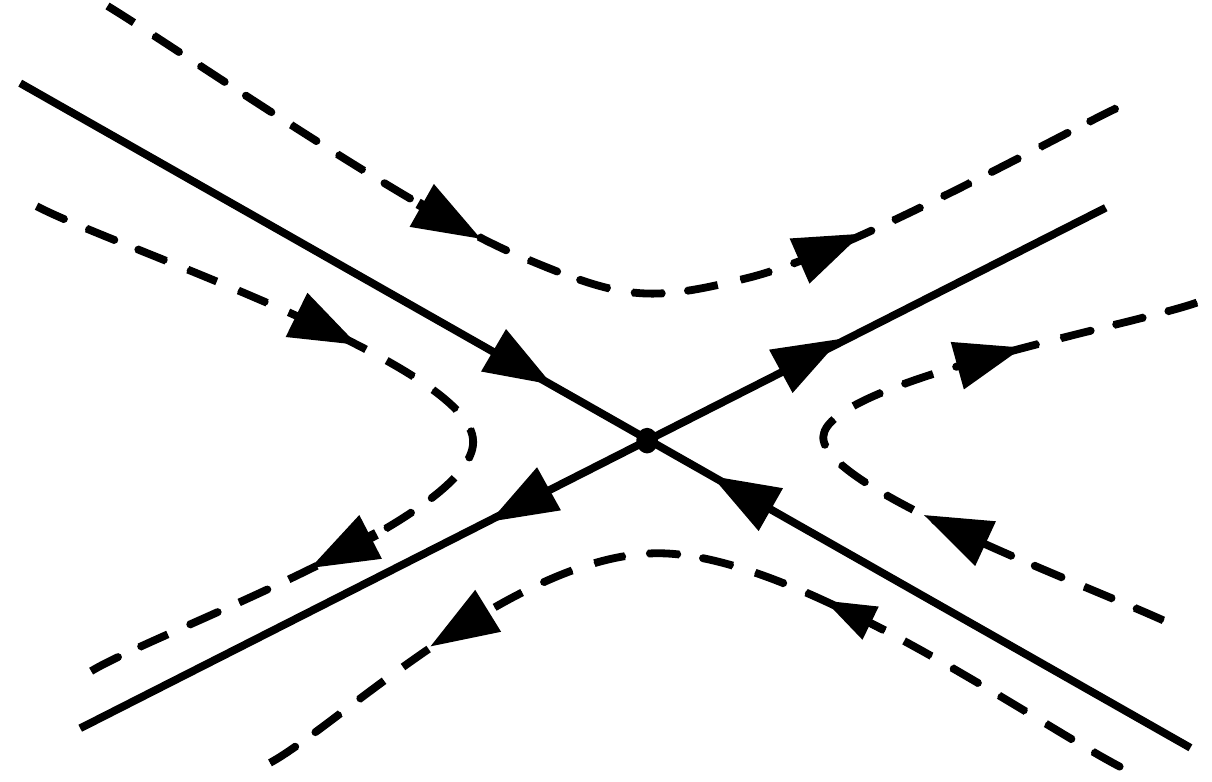}
		\caption{\footnotesize saddle}
		\label{f:saddle}		
	\end{subfigure}
	\quad
	\caption{\footnotesize Poincar\'e's classification of zeros of a vector field (equilibrium or fixed points) on a plane. (a) Center is always stable with oscillatory motion nearby, (b,c) nodes and foci (or spirals) can be stable or unstable and (d) saddles are unstable except in one direction.}
	\label{f:zeroes-classification}
\end{figure}

\begin{figure}[h] 
\center
\includegraphics[width=4cm]{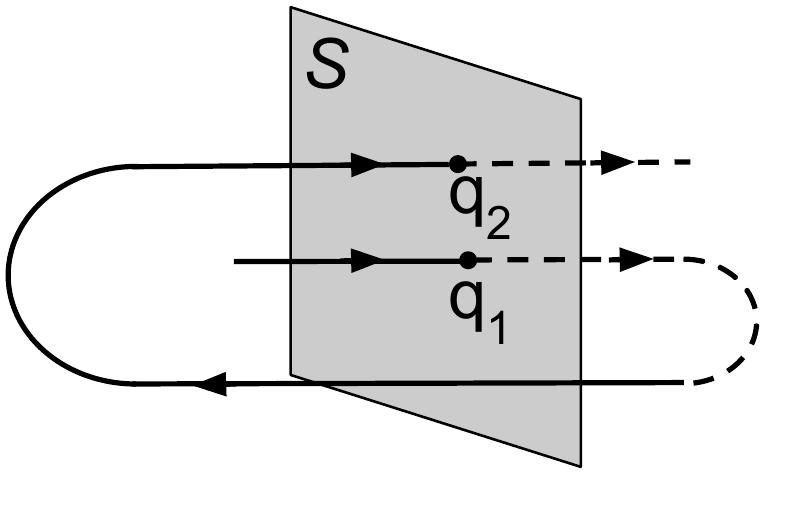}
\caption{\footnotesize A Poincare surface $S$ transversal to a trajectory is shown. The trajectory through $q_1$ on $S$ intersects $S$ again at $q_2$. The map taking $q_1$ to $q_2$ is called Poincar\'e's first return map.}
\label{f:poincare-return-map}
\end{figure}

\begin{figure}[h] 
\center
\includegraphics[width=4cm]{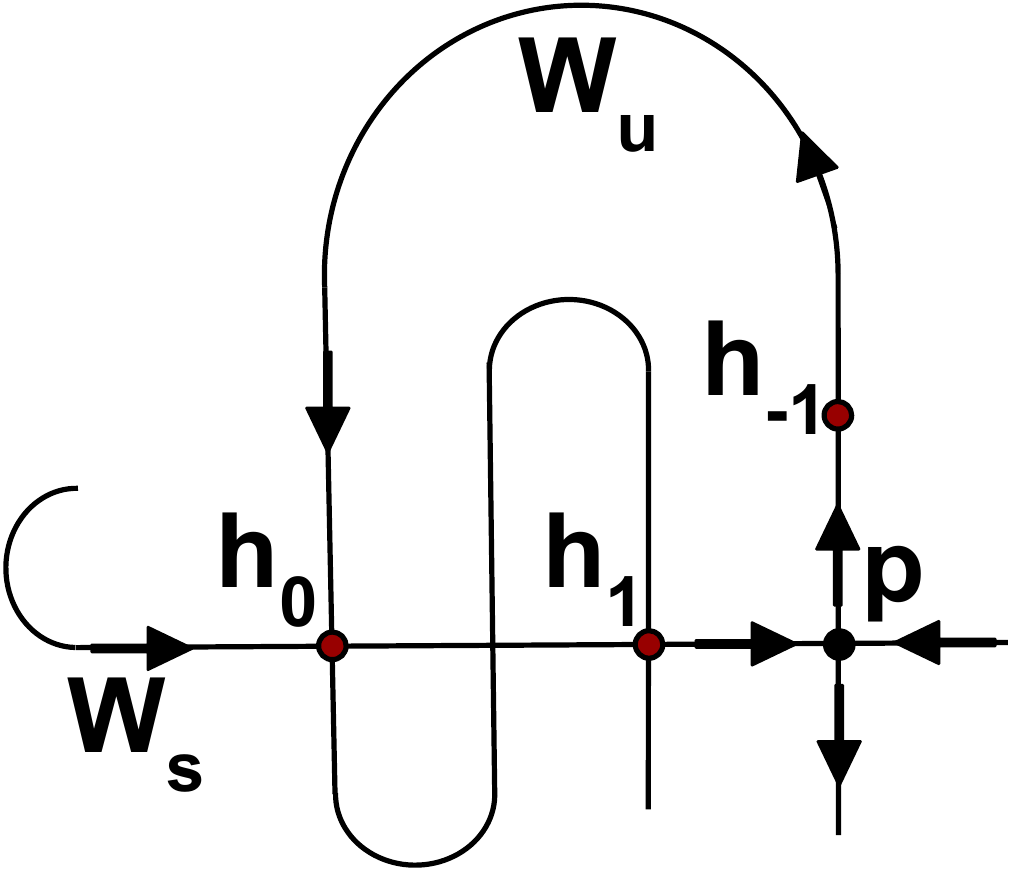}
\caption{\footnotesize The saddle point $p$ and its stable and unstable spaces $W_s$ and $W_u$ are shown on a Poincar\'e surface through $p$. The points at which $W_s$ and $W_u$ intersect are called homoclinic points, e.g., $h_0,$ $h_1$ and $h_{-1}$. Points on $W_s$ (or $W_u$) remain on $W_s$ (or $W_u$) under forward and backward iterations of the return map. Thus, the forward and backward images of a homoclinic point under the return map are also homoclinic points. In the figure $h_0$ is a homoclinic point whose image is $h_1$ on the segment $[h_0,p]$ of $W_s$. Thus, $W_u$ must fold back to intersect $W_s$ at $h_1$. Similarly, if $h_{-1}$ is the backward image of $h_0$ on $W_u$, then $W_s$ must fold back to intersect $W_u$ at $h_{-1}$. Further iterations produce an infinite number of homoclinic points accumulating at $p$. The first example of a homoclinic tangle was discovered by Poincar\'e in the restricted 3-body problem and is a signature of its chaotic nature.}
\label{f:homoclinic-points}
\end{figure}

The French mathematician/physicist/engineer Henri Poincar\'e began by developing a qualitative theory of differential equations from a global geometric viewpoint of the dynamics on phase space. This included a classification of the types of equilibria (zeros of vector fields) on the phase plane (nodes, saddles, foci and centers, see Fig.~\ref{f:zeroes-classification}). His 1890 memoir on the three-body problem was the prize-winning entry in King Oscar II's $60^{\rm th}$ birthday competition (for a detailed account see \cite{barrow-green-poincare-three-body}). He proved the divergence of series solutions for the 3-body problem developed by Delaunay, Hugo Gyld\'en and Lindstedt (in many cases) and covergence of Hill's infinite determinants. To investigate the stability of 3-body motions, Poincar\'e defined his `surfaces of section' and a discrete-time dynamics via the `return map' (see Fig.~\ref{f:poincare-return-map}). A Poincar\'e surface $S$ is a two-dimensional surface in phase space transversal to trajectories. The first return map takes a point $q_1$ on $S$ to $q_2$, which is the next intersection of the trajectory through $q_1$ with $S$. Given a saddle point $p$ on a surface $S$, he defined its stable and unstable spaces $W_s$ and $W_u$ as points on $S$ that tend to $p$ upon repeated forward or backward applications of the return map (see Fig.~\ref{f:homoclinic-points}). He initially assumed that $W_s$ and $W_u$ on a surface could not intersect and used this to argue that the solar system is stable. This assumption turned out to be false, as he discovered with the help of Lars Phragm\'en. In fact, $W_s$ and $W_u$ can intersect transversally on a surface at a homoclinic point\footnote{Homoclinic refers to the property of being `inclined' both forward and backward in time to the same point.} if the state space of the underlying continuous dynamics is at least three-dimensional. What is more, he showed that if there is one homoclinic point, then there must be infinitely many accumulating at $p$. Moreover, $W_s$ and $W_u$ fold and intersect in a very complicated `homoclinic tangle' in the vicinity of $p$. This was the first example of what we now call chaos. Chaos is usually manifested via an extreme sensitivity to initial conditions (exponentially diverging trajectories with nearby initial conditions).

When two gravitating point masses collide, their relative speed diverges and solutions to the equations of motion become singular at the collision time $t_c$. More generally, a singularity occurs when either a position or velocity diverges in finite time. The Frenchman Paul Painlev\'e (1895) showed that binary and triple collisions are the only possible singularities in the three-body problem. However, he conjectured that non-collisional singularities (e.g. where the separation between a pair of bodies goes to infinity in finite time) are possible for four or more bodies. It took nearly a century for this conjecture to be proven, culminating in the work of Donald Saari and Zhihong Xia (1992) and Joseph Gerver (1991) who found explicit examples of non-collisional singularities in the $5$-body and $3n$-body problems for $n$ sufficiently large \cite{saari}. In Xia's example, a particle oscillates with ever growing frequency and amplitude between two pairs of tight binaries. The separation between the binaries diverges in finite time, as does the velocity of the oscillating particle.

The Italian mathematician Tulio Levi-Civita (1901) attempted to avoid singularities and thereby `regularize' collisions in the three-body problem by a change of variables in the differential equations. For example, the ODE for the one-dimensional Kepler problem $\ddot x = - k/x^2$ is singular at the collision point $x=0$. This singularity can be regularized\footnote{Solutions which could be smoothly extended beyond collision time (e.g., the bodies elastically collide) were called regularizable. Those that could not were said to have an essential or transcendent singularity at the collision.} by introducing a new coordinate $x = u^2$ and a reparametrized time $ds = dt/u^2$, which satisfy the nonsingular oscillator equation $u''(s) = E u/2$ with conserved energy $E = (2 \dot u^2 - k)/u^2$. Such regularizations could shed light on near-collisional trajectories (`near misses') provided the differential equations remain physically valid\footnote{Note that the point particle approximation to the equations for celestial bodies of non-zero size breaks down due to tidal effects when the bodies get very close}. 

The Finnish mathematician Karl Sundman (1912) began by showing that binary collisional singularities in the 3-body problem could be regularized by a repararmetrization of time, $s = |t_1-t|^{1/3}$ where $t_1$ is the the binary collision time \cite{siegel-moser}. He used this to find a {\it convergent} series representation (in powers of $s$) of the general solution of the 3-body problem in the absence of triple collisions\footnote{Sundman showed that for non-zero angular momentum, there are no triple collisions in the three-body problem.}. The possibility of such a convergent series had been anticipated by Karl Weierstrass in proposing the 3-body problem for King Oscar's 60th birthday competition. However, Sundman's series converges exceptionally slowly and has not been of much practical or qualitative use.

\begin{figure}[h] 
\center
\includegraphics[width=5cm]{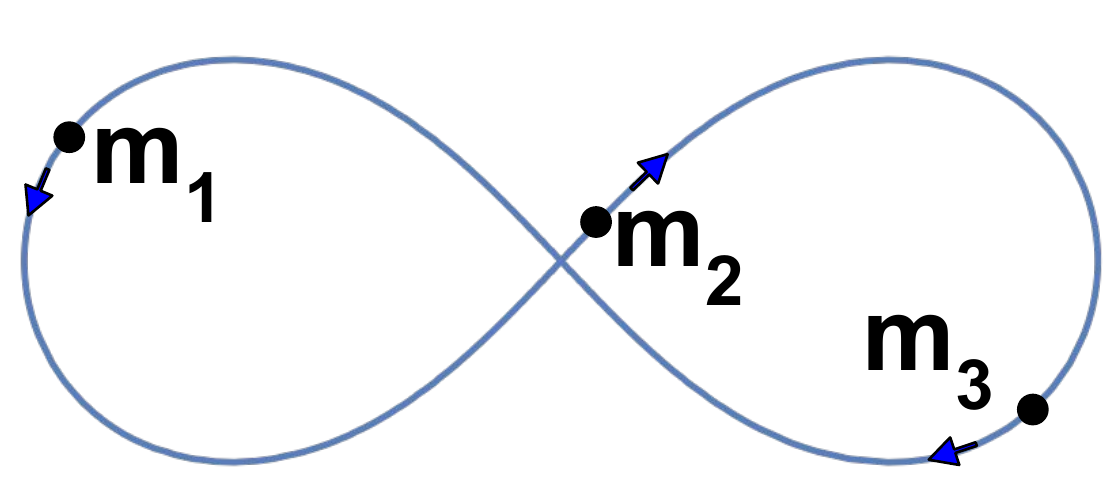}
\caption{\footnotesize Equal-mass zero-angular momentum figure-8 choreography solution to the 3-body problem. A choreography is a periodic solution where all masses traverse the same orbit separated equally in time.}
\label{f:figure-8}
\end{figure}

The advent of computers in the $20^{\rm th}$ century allowed numerical investigations into the 3-body (and more generally the $n$-body) problem. Such numerical simulations have made possible the accurate placement of satellites in near-Earth orbits as well as our missions to the Moon, Mars and the outer planets. They have also facilitated theoretical explorations of the three-body problem including chaotic behavior, the possibility for ejection of one body at high velocity (seen in hypervelocity stars \cite{hypervelocity-stars}) and quite remarkably, the discovery of new periodic solutions. For instance, in 1993, Chris Moore discovered the zero angular momentum figure-8 `choreography' solution. It is a stable periodic solution with bodies of equal masses chasing each other on an $\infty$-shaped trajectory while separated equally in time (see Fig.~\ref{f:figure-8}). Alain Chenciner and Richard Montgomery \cite{montgomery-notices-ams} proved its existence using an elegant geometric reformulation of Newtonian dynamics that relies on the variational principle of Euler and Maupertuis.

%-------------
\section{Geometrization of mechanics}
%------------

Fermat's principle in optics states that light rays extremize the optical path length $\int n(\bfr(\tau)) \: d\tau$ where $n(\bfr)$ is the (position dependent) refractive index and $\tau$ a parameter along the path\footnote{The optical path length $\int n(\bfr) \, d\tau$ is proportional to $\int d\tau/\la$, which is the geometric length in units of the local wavelength $\la(\bfr) = c/n(\bfr) \nu$. Here, $c$ is the speed of light in vacuum and $\nu$ the constant frequency.}. The variational principle of Euler and Maupertuis (1744) is a mechanical analogue of Fermat's principle \cite{lanczos}. It states that the curve that extremizes the abbreviated action $\int_{\bfq_1}^{\bfq_2} {\bf p}\cdot d{\bf q}$ holding energy $E$ and the end-points $\bfq_1$ and $\bfq_2$ fixed has the same shape as the Newtonian trajectory. By contrast, Hamilton's principle of extremal action (1835) states that a trajectory going from $\bfq_1$ at time $t_1$ to $\bfq_2$ at time $t_2$ is a curve that extremizes the action\footnote{The action is the integral of the Lagrangian $S = \int_{t_1}^{t_2} L(\bfq,\dot \bfq) \: dt$. Typically, $L = T - V$ is the difference between kinetic and potential energies.}.

It is well-known that the trajectory of a free particle (i.e., subject to no forces) moving on a plane is a straight line. Similarly, trajectories of a free particle  moving on the surface of a sphere are great circles. More generally, trajectories of a free particle moving on a curved space (Riemannian manifold $M$) are geodesics (curves that extremize length). Precisely, for a mechanical system with configuration space $M$ and Lagrangian $L = \half m_{ij}(\bfq) \dot q^i \dot q^j$, Lagrange's equations $\DD{p_i}{t} = \dd{L}{q^i}$ are equivalent to the geodesic equations with respect to the `kinetic metric' $m_{ij}$ on $M$\footnote{A metric $m_{ij}$ on an $n$-dimensional configuration space $M$ is an $n \times n$ matrix at each point $\bfq \in M$ that determines the square of the distance ($ds^2 = \sum_{i,j = 1}^n m_{ij} dq^i dq^j$) from $\bfq$ to a nearby point $\bfq + d \bfq$. We often suppress the summation symbol and follow the convention that repeated indices are summed from $1$ to $n$.}:
	\beq
	 m_{ij} \: \ddot q^j(t) = - \half \left(m_{ji,k} + m_{ki,j} - m_{jk,i} \right) \dot q^j(t) \: \dot q^k(t).
	 \label{e:Lagrange-eqns-kin-metric-and-V}
	 \eeq
Here, $m_{ij,k} = \pdr m_{ij}/\pdr q^k$ and $p_i = \dd{L}{\dot q^i} = m_{ij}\dot q^j$ is the momentum conjugate to coordinate $q^i$. For instance, the kinetic metric ($m_{rr} = m$, $m_{\tht \tht} = m r^2$, $m_{r \tht} = m_{\tht r} = 0$) for a free particle moving on a plane may be read off from the Lagrangian $L = \half m (\dot r^2 + r^2 \dot \tht^2)$ in polar coordinates, and the geodesic equations shown to reduce to Lagrange's equations of motion $\ddot r = r \dot \tht^2$ and $d(m r^2 \dot \tht)/dt = 0$.

Remarkably, the correspondence between trajectories and geodesics continues to hold even in the presence of conservative forces derived from a potential $V$. Indeed, trajectories of the Lagrangian $L = T - V = \half m_{ij}(\bfq) \dot q^i \dot q^j - V(\bfq)$ are {\it reparametrized}\footnote{The shapes of trajectories and geodesics coincide but the Newtonian time along trajectories is not the same as the arc-length parameter along geodesics.} geodesics of the Jacobi-Maupertuis (JM) metric $g_{ij} = (E- V(\bfq)) m_{ij}(\bfq)$ on $M$ where $E = T + V$ is the energy. This geometric formulation of the Euler-Maupertuis principle (due to Jacobi) follows from the observation that the square of the metric line element
	\beq
	ds^2 = g_{ij} dq^i dq^j = (E-V) m_{ij} dq^i dq^j = \half m_{kl} \fr{dq^k}{dt} \fr{dq^l}{dt} m_{ij} dq^i dq^j = \half \left( m_{ij} \dot q^i dq^j \right)^2 = \ov{2} (\bfp \cdot d\bfq)^2,
	\eeq
so that the extremization of $\int \bfp \cdot d\bfq$ is equivalent to the extremization of arc length $\int ds$. Loosely, the potential $V(\bfq)$ on the configuration space plays the role of an inhomogeneous refractive index. Though trajectories and geodesics are the same curves, the Newtonian time $t$ along trajectories is in general different from the arc-length parameter $s$ along geodesics. They are related by $\DD{s}{t} = \sqrt{2} (E-V)$ \cite{govind-himalaya}.

\begin{figure}	
	\centering
	\begin{subfigure}[t]{5cm}
		\centering
		\includegraphics[width=4cm]{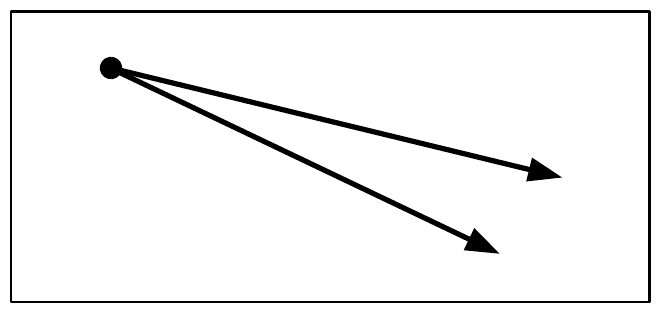}
		\caption{\footnotesize Nearby geodesics on a plane ($K = 0$) separate linearly.}
		\label{f:planar-geodesics}		
	\end{subfigure}
\quad
	\begin{subfigure}[t]{5cm}
		\centering
		\includegraphics[width=2.3cm]{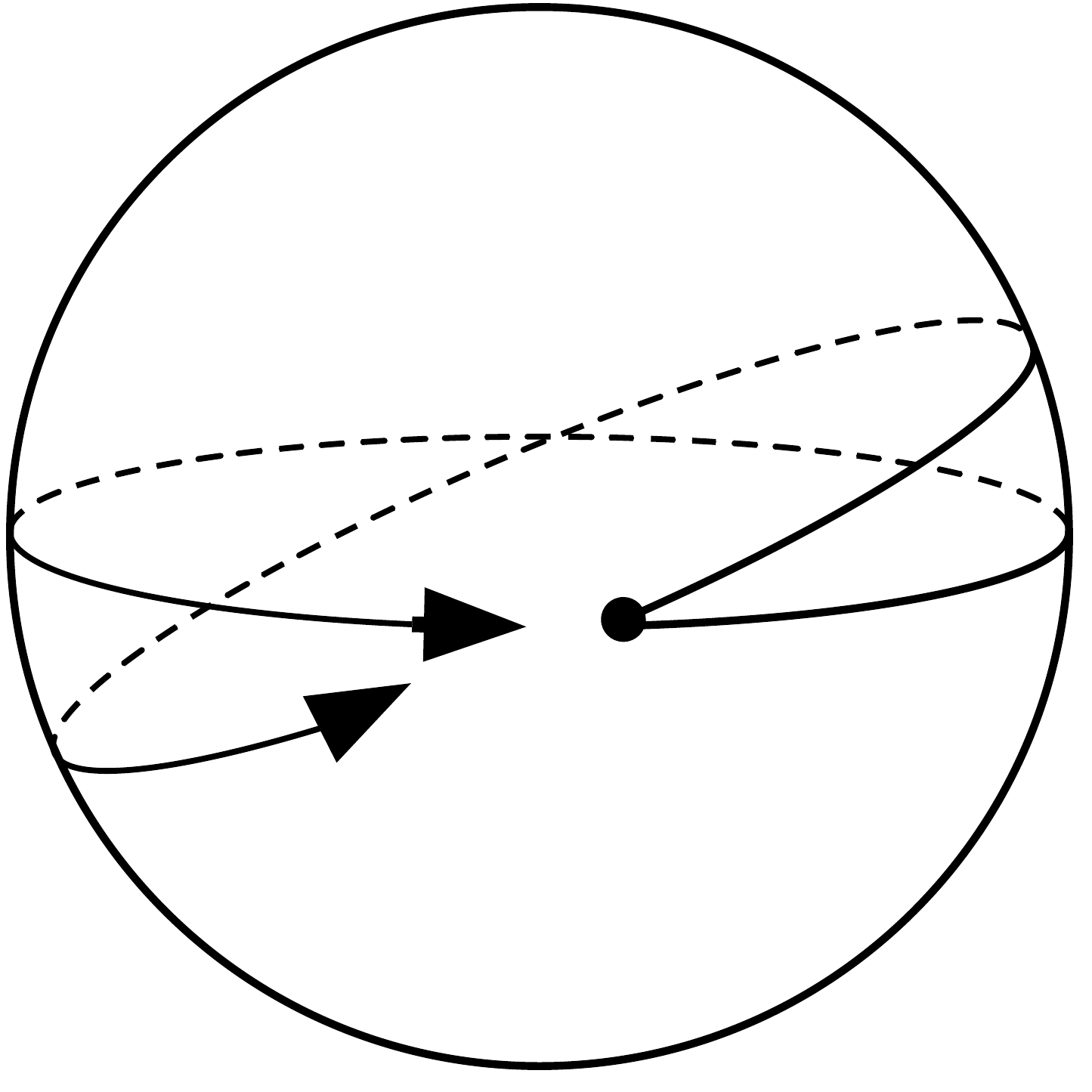}
		\caption{\footnotesize Distance between neighboring geodesics on a sphere ($K > 0$) oscillates.}
		\label{f:spherical-geodesics}		
	\end{subfigure}		
\quad
	\begin{subfigure}[t]{5cm}
		\centering
		\includegraphics[width=4cm]{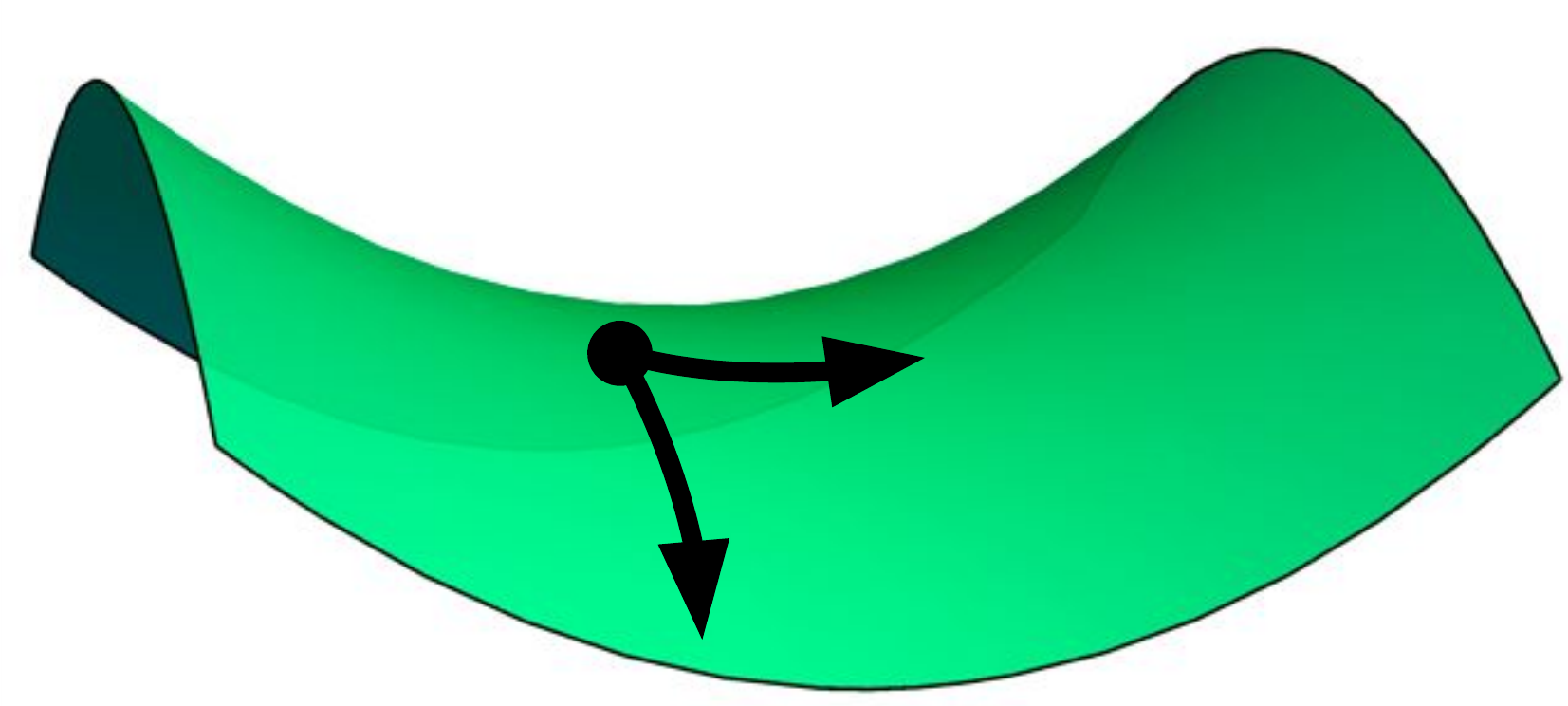}
		\caption{\footnotesize Geodesics on a hyperbolic surface ($K < 0$) deviate exponentially}
		\label{f:hyperbolic-geodesics}		
	\end{subfigure}
	\caption{\footnotesize Local behavior of nearby geodesics on a surface depends on the sign of its Gaussian curvature $K$.}
	\label{f:geodesic-separation}
\end{figure}

This geometric reformulation of classical dynamics allows us to assign a local curvature to points on the configuration space. For instance, the Gaussian curvature $K$ of a surface at a point (see Box 11) measures how nearby geodesics behave (see Fig. \ref{f:geodesic-separation}), they oscillate if $K > 0$ (as on a sphere), diverge exponentially if $K < 0$ (as on a hyperboloid) and linearly separate if $K = 0$ (as on a plane). Thus, the curvature of the Jacobi-Maupertuis metric defined above furnishes information on the stability of trajectories. Negativity of curvature leads to sensitive dependence on initial conditions and can be a source of chaos. 

\begin{center}
	\begin{mdframed}
{\bf Box 11: Gaussian curvature:} Given a point $p$ on a surface $S$ embedded in three dimensions, a normal plane through $p$ is one that is orthogonal to the tangent plane at $p$. Each normal plane intersects $S$ along a curve whose best quadratic approximation at $p$ is called its osculating circle. The principal radii of curvature $R_{1,2}$ at $p$ are the maximum and minimum radii of osculating circles through $p$. The Gaussian curvature $K(p)$ is defined as $1/R_1 R_2$ and is taken positive if the centers of the corresponding osculating circles lie on the same side of $S$ and negative otherwise.
	\end{mdframed}
\end{center}

In the planar Kepler problem, the Hamiltonian (\ref{e:energy-kepler-cm-frame}) in the CM frame is
	\beq
	H = \fr{p_x^2+p_y^2}{2m} - \fr{\al}{r} \quad \text{where} \quad \al = GMm > 0 \;\;\text{and} \;\; r^2 = x^2 + y^2.
	\eeq
The corresponding JM metric line element in polar coordinates is $ds^2 = m\left(E+\fr{\al}{r}\right)\left(dr^2+r^2d\theta^2\right)$. Its Gaussian curvature $K = -E \al/2m(\al + Er)^3$ has a sign opposite to that of energy everywhere. This reflects the divergence of nearby hyperbolic orbits and oscillation of nearby elliptical orbits. Despite negativity of curvature and the consequent sensitivity to initial conditions, hyperbolic orbits in the Kepler problem are not chaotic: particles simply fly off to infinity and trajectories are quite regular. On the other hand, negativity of curvature without any scope for escape can lead to chaos. This happens with geodesic motion on a compact Riemann surface\footnote{ A compact Riemann surface is a closed, oriented and bounded surface such as a sphere, a torus or the surface of a pretzel. The genus of such a surface is the number of handles: zero for a sphere, one for a torus and two or more for higher handle-bodies. Riemann surfaces with genus two or more admit metrics with constant negative curvature.} with constant negative curvature: most trajectories are very irregular.

%-------------
\section{Geometric approach to the planar 3-body problem}
%-------------

\begin{figure}	
	\centering
		\begin{subfigure}[t]{3in}
		\centering
		\includegraphics[width=5cm]{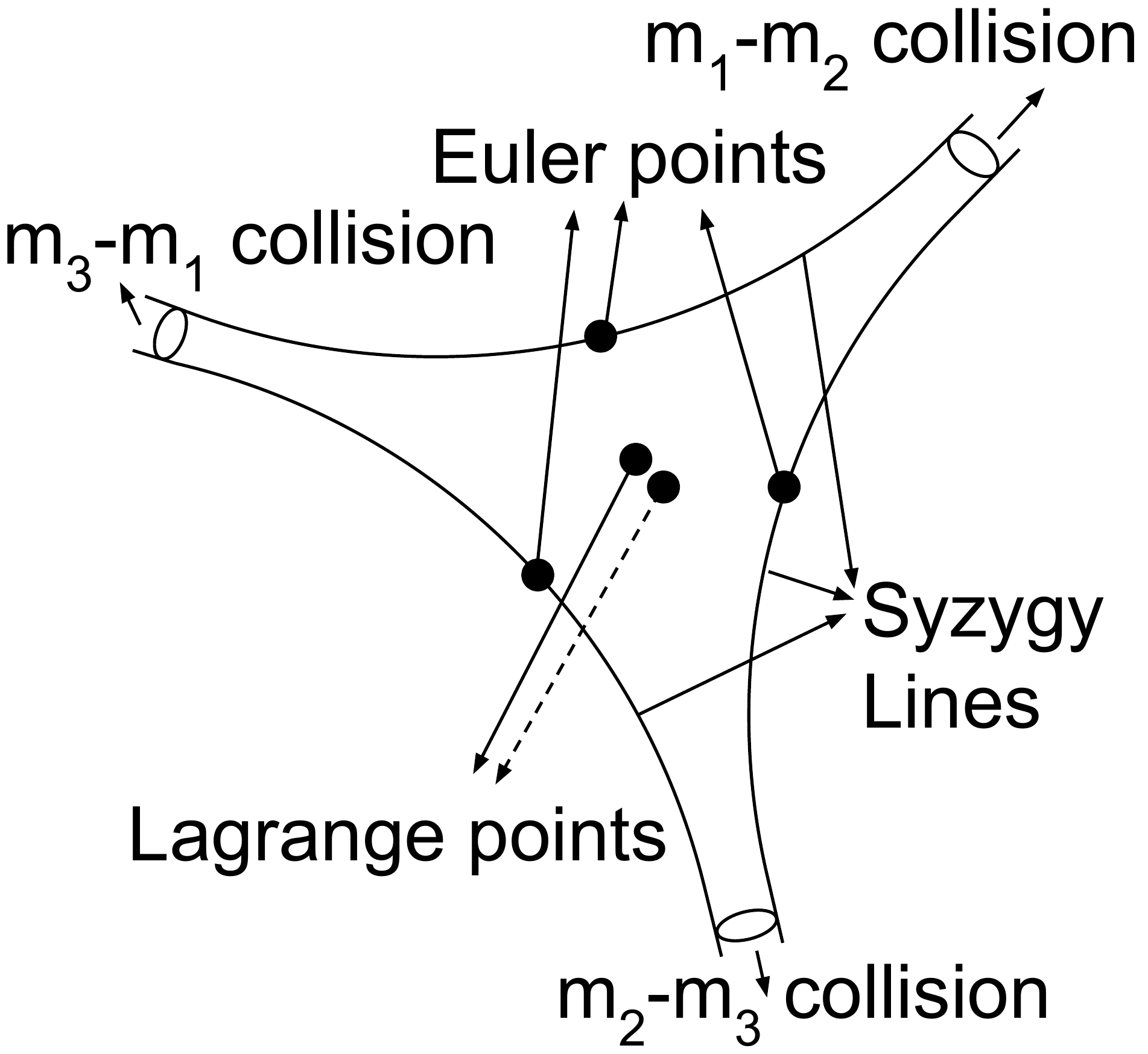}
		\caption{\footnotesize The negatively curved `pair of pants' metric on the shape sphere $\mS^2$.}
		\label{f:horn-shape-sphere}		
	\end{subfigure}	
\quad
	\begin{subfigure}[t]{3in}
		\centering
		\includegraphics[width=5cm]{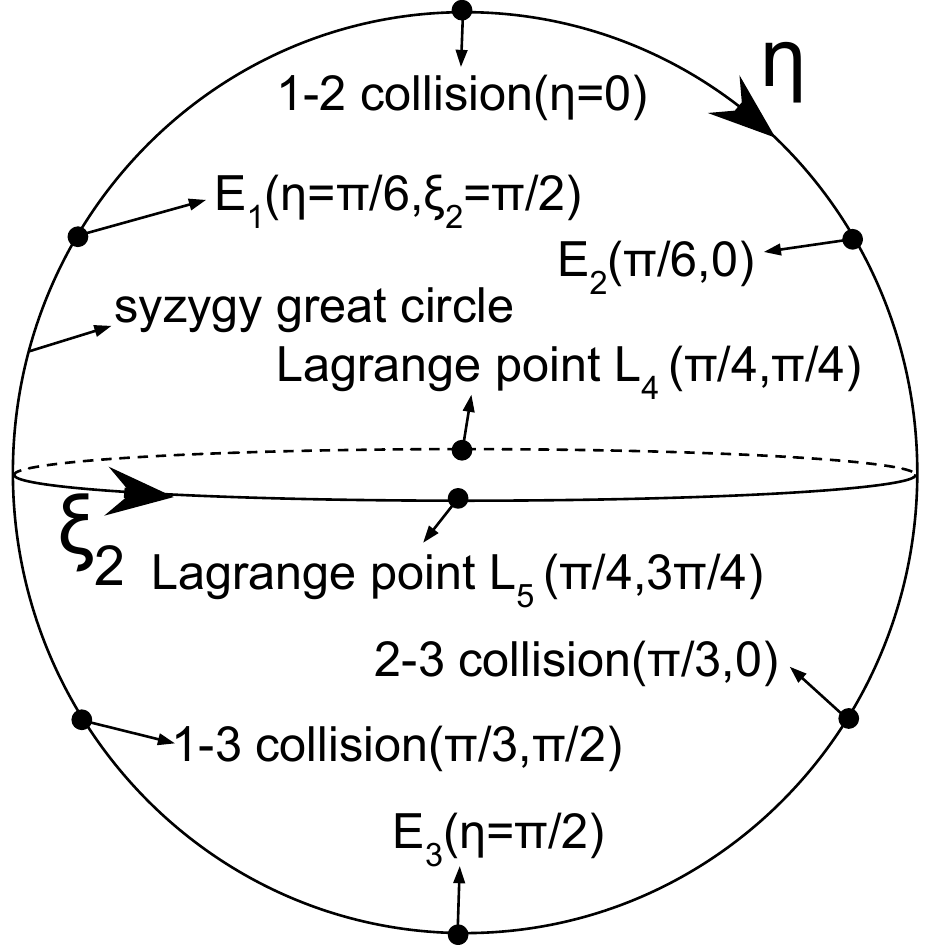}
		\caption{\footnotesize Locations of Lagrange, Euler and collision points on a geometrically {\it unfaithful} depiction of the shape sphere $\mS^2$. 
The negative curvature of $\mS^2$ is indicated in Fig.~\ref{f:horn-shape-sphere}. Syzygies are instantaneous configurations where the three bodies are collinear (eclipses).}
		\label{f:round-shape-sphere}	
	\end{subfigure}	\caption{\footnotesize `Pair of pants' metric on shape sphere and Lagrange, Euler and collision points.}
	\label{f:shape-sphere}
\end{figure}

We now sketch how the above geometrical framework may be usefully applied to the three-body problem. The configuration space of the planar 3-body problem is the space of triangles on the plane with masses at the vertices. It may be identified with six-dimensional Euclidean space ($\mathbb{R}^6$) with the three planar Jacobi vectors $\bfJ_{1,2,3}$ (see (\ref{e:jacobi-coord}) and Fig.~\ref{f:jacobi-coords}) furnishing coordinates on it. A simultaneous translation of the position vectors of all three bodies $\bfr_{1,2,3} \mapsto \bfr_{1,2,3} + \bfr_0$ is a symmetry of the Hamiltonian $H = T+V$ of Eqs. (\ref{e:jacobi-coord-ke-mom-inertia},\ref{e:jacobi-coord-potential}) and of the Jacobi-Maupertuis metric 
	\beq
	\label{e:jm-metric-in-jacobi-coordinates-on-c3}
	ds^2 = \left( E - V(\bfJ_1, \bfJ_2)  \right) \sum_{a=1}^3 M_a \: |d\bfJ_a|^2. 
	\eeq
This is encoded in the cyclicity of $\bfJ_3$. Quotienting by translations allows us to define a center of mass configuration space $\mathbb{R}^4$ (the space of centered triangles on the plane with masses at the vertices) with its quotient JM metric. Similarly, rotations $\bfJ_a \to \colvec{2}{\cos \tht & -\sin \tht}{\sin \tht & \cos \tht} \bfJ_a$ for $a = 1,2,3$ are a symmetry of the metric, corresponding to rigid rotations of a triangle about a vertical axis through the CM. The quotient of $\mR^4$ by such rotations is the {\it shape space} $\mR^3$, which is the space of congruence classes of centered oriented triangles on the plane. Translations and rotations are symmetries of any central inter-particle potential, so the dynamics of the three-body problem in any such potential admits a consistent reduction to geodesic dynamics on the shape space $\mR^3$. Interestingly, for an {\it inverse-square} potential (as opposed to the Newtonian `$1/r$' potential)
	\beq
	V = -\sum_{a < b} \fr{G m_a m_b}{|\bfr_a - \bfr_b|^2} = -\fr{G m_1 m_2}{|\bfJ_1|^2} - \fr{G m_2 m_3}{|\bfJ_2 - \mu_1 \bfJ_1|^2} - \fr{G m_3 m_1}{|\bfJ_2+\mu_2 \bfJ_1|^2} \quad \text{with} \quad
	 \mu_{1,2}= \frac{m_{1,2}}{m_1 + m_2},
	\eeq
the zero-energy JM metric (\ref{e:jm-metric-in-jacobi-coordinates-on-c3}) is also invariant under the scale transformation $\bfJ_a \to \la \bfJ_a$ for $a = 1,2$ and $3$ (see Box 12 for more on the inverse-square potential and for why the zero-energy case is particularly interesting). This allows us to further quotient the shape space $\mR^3$ by scaling to get the shape sphere $\mS^2$, which is the space of similarity classes of centered oriented triangles on the plane\footnote{Though scaling is not a symmetry for the Newtonian gravitational potential, it is still useful to project the motion onto the shape sphere.}. Note that collision configurations are omitted from the configuration space and its quotients. Thus, the shape sphere is topologically a $2$-sphere with the three binary collision points removed. In fact, with the JM metric, the shape sphere looks like a `pair of pants' (see Fig.~\ref{f:horn-shape-sphere}).

For equal masses and $E=0$, the quotient JM metric on the shape sphere may be put in the form 
	\beq
	\label{e:jm-metric-zero-energy-shape-sphere}
	ds^2 = Gm^3 h(\eta,\xi_2) \left(d\eta^2+\sin^2 2\eta \;d\xi_2^2\right).
	\eeq
Here, $0 \le 2 \eta \le \pi$ and $0 \le 2 \xi_2 \le 2 \pi$ are polar and azimuthal angles on the shape sphere $\mS^2$ (see Fig.~\ref{f:round-shape-sphere}). The function $h$ is invariant under the above translations, rotations and scalings and therefore a function on $\mS^2$. It may be written as $v_1 + v_2 + v_3$ where $v_1 = I_{\rm CM}/(m |\bfr_2 - \bfr_3|^2)$ etc., are proportional to the inter-particle  potentials \cite{govind-himalaya}. As shown in Fig.~\ref{f:horn-shape-sphere}, the shape sphere has three cylindrical horns that point toward the three collision points, which lie at an infinite geodesic distance. Moreover, this equal-mass, zero-energy JM metric (\ref{e:jm-metric-zero-energy-shape-sphere}) has negative Gaussian curvature everywhere except at the Lagrange and collision points where it vanishes. This negativity of curvature implies geodesic instability (nearby geodesics deviate exponentially) as well as the uniqueness of geodesic representatives in each `free' homotopy class, when they exist. The latter property was used by Montgomery \cite{montgomery-notices-ams} to establish uniqueness of the `figure-8' solution (up to translation, rotation and scaling) for the inverse-square potential. The negativity of curvature on the shape sphere for equal masses extends to negativity of scalar curvature\footnote{Scalar curvature is an average of the Gaussian curvatures in the various tangent planes through a point} on the CM configuration space for both the inverse-square and Newtonian gravitational potentials \cite{govind-himalaya}. This could help to explain instabilities and chaos in the three-body problem.

%-----------------

\begin{center}
	\begin{mdframed}
{\bf Box 12:} {\bf The inverse-square potential} is somewhat simpler than the Newtonian one due to the behavior of the Hamiltonian $H = \sum_a \bfp_a^2/2m_a - \sum_{a < b} G m_a m_b/|\bfr_a -\bfr_b|^2$ under scale transformations $\bfr_a \to \la \bfr_a$ and $\bfp_a \to \la^{-1} \bfp_a$:  $H(\la \bfr, \la^{-1} \bfp) = \la^{-2} H(\bfr, \bfp)$ \cite{Rajeev}. The infinitesimal version ($\la \approx 1$) of this transformation is generated by the dilatation operator $D = \sum_a \bfr_a \cdot \bfp_a$ via Poisson brackets $\{\bfr_a, D \} = \bfr_a$ and $\{\bfp_a, D \} = - \bfp_a$. Here, the Poisson bracket between coordinates and momenta are $\{ r_{ai}, p_{bj} \} = \del_{ab} \del_{ij}$ where $a,b$ label particles and $i,j$ label Cartesian components. In terms of Poisson brackets, time evolution of any quantity $f$ is given by $\dot f = \{ f, H \}$. It follows that $\dot D = \{ D, H \} = 2 H$, so scaling is a symmetry of the Hamiltonian (and $D$ is conserved) only when the energy vanishes. To examine long-time behavior we consider the evolution of the moment of inertia in the CM frame $I_{\rm CM} = \sum_a m_a \bfr_a^2$ whose time derivative may be expressed as $\dot I = 2D$. This leads to the Lagrange-Jacobi identity $\ddot I = \{\dot I, H \} = \{2D, H \} = 4 E$ or $I = I(0) + \dot I(0) \: t + 2E \: t^2$. Hence when $E > 0$, $I \to \infty$ as $t \to \infty$ so that bodies fly apart asymptotically. Similarly, when $E < 0$ they suffer a triple collision. When $E = 0$, the sign of $\dot I(0)$ controls asymptotic behavior leaving open the special case when $E = 0$ and $\dot I(0) = 0$. By contrast, for the Newtonian potential, the Hamiltonian transforms as $H(\la^{-2/3} \bfr, \la^{1/3} \bfp) = \la^{2/3} H(\bfr, \bfp)$ leading to the Lagrange-Jacobi identity $\ddot I = 4E - 2V$. This is however not adequate to determine the long-time behavior of $I$ when $E < 0$.
	\end{mdframed}
\end{center}
%-----------------

%================================


\begin{thebibliography}{99}

\bibitem{laskar} Laskar, J., {\it Is the Solar System stable?} Progress in Mathematical Physics, {\bf 66}, 239-270 (2013).

\bibitem{gutzwiller-book} Gutzwiller, M. C., {\it Chaos in Classical and Quantum mechanics}, Springer-Verlag, New York (1990).

\bibitem{goldstein} Goldstein, H.,  Poole, C. P.,  and Safko, J. L.,  {\it Classical Mechanics}, 3rd Ed., Pearson Education (2011).

\bibitem{hand-finch} Hand, L. N. and Finch, J. D., {\it Analytical Mechanics}, Cambridge Univ. Press (1998).


\bibitem{Rajeev} Rajeev, S. G., {\it Advanced Mechanics: From Euler's Determinism to Arnold's Chaos}, Oxford University Press, Oxford (2013).


\bibitem{diacu-holmes} Diacu F. and Holmes P., {\it Celestial Encounters: The Origins of Chaos and Stability}, Princeton University Press, New Jersey (1996).

\bibitem{musielak-quarles} Musielak, Z. E. and Quarles B., {\it The three-body problem}, Reports on Progress in Physics, {\bf 77}, 6, 065901 (2014), arXiv:1508.02312.


\bibitem{barrow-green-poincare-three-body} Barrow-Green, J., {\it Poincar\'e and the Three Body Problem}, Amer. Math. Soc., Providence, Rhode Island (1997).

\bibitem{whittaker} Whittaker, E. T., {\it A treatise on the analytical dynamics of particles \& rigid bodies}, 2nd Ed., Cambridge University Press, Cambridge (1917), Chapt. XIV and page 283.

\bibitem{symon} Symon, K. R., {\it Mechanics}, 3rd Ed., Addison Wesley, Philippines (1971).

\bibitem{bodenmann-lunar-battle} Bodenmann, S., {\it The 18th-century battle over lunar motion}, Physics Today, {\bf 63}(1), 27 (2010). % doi: 10.1063/1.3293410

\bibitem{mukunda-hamilton} Mukunda, N., {\it Sir William Rowan Hamilton}, Resonance, {\bf 21} (6), 493 (2016).


\bibitem{gutzwiller-three-body} Gutzwiller, M. C., {\it Moon-Earth-Sun: The oldest three-body problem}, Reviews of Modern Physics, {\bf 70}, 589 (1998).

\bibitem{saari} Saari, D. G. and Xia, Z., {\it Off to infinity in finite time}, Notices of the AMS, {\bf 42}, 538 (1993). 


\bibitem{siegel-moser} Siegel, C.L. and Moser, J.K., {\it Lectures on Celestial Mechanics}, Springer-Verlag, Berlin (1971), page 31.

\bibitem{hypervelocity-stars} Brown, W. R., {\it Hypervelocity Stars in the Milky Way}, Physics Today, {\bf 69}(6), 52 (2016).

\bibitem{montgomery-notices-ams} Montgomery, R., {\it A new solution to the three-body problem}, Notices of the AMS, {\bf 48}(5), 471 (2001).

\bibitem{lanczos} Lanczos, C., {\it The variational principles of mechanics}, 4th Ed., Dover, New York (1970),  page 139.

\bibitem{govind-himalaya} Krishnaswami, G. S. and Senapati, H., {\it Curvature and geodesic instabilities in a geometrical approach to the planar three-body problem}, J. Math. Phys., {\bf 57}, 102901 (2016), arXiv:1606.05091.





\end{thebibliography}
\end{document}